\documentclass[journal, 11pt, draftcls, onecolumn]{IEEEtran}

\usepackage{cite}
\usepackage{eqparbox}
\usepackage{stfloats}
\usepackage{url}
\usepackage[caption=false,font=footnotesize]{subfig}
\usepackage{graphicx}
\usepackage{epstopdf}
\usepackage[cmex10]{amsmath}
\usepackage[]{units}
\usepackage[pagewise]{lineno}
\usepackage{array}

\newtheorem{theorem}{Theorem}
\newtheorem{remark}{Remark}

\usepackage{color}

\begin{document}
\linenumbers
\newcommand{\lsc}[1]{\textcolor{black}{#1}}
\newcommand{\yq}[1]{\textcolor{black}{#1}}
%
\title{Asynchronous Convolutional-Coded Physical-Layer Network Coding}
%
%
%

\author{Qing~Yang,~\IEEEmembership{Student Member,~IEEE,} and
        Soung~Chang~Liew,~\IEEEmembership{Fellow,~IEEE}%
\thanks{The authors are with the Department of Information Engineering, The Chinese University of Hong Kong, Shatin, New Territories, Hong Kong (email: \{yq010, soung\}@ie.cuhk.edu.hk).}

\thanks{This work is partially supported by AoE grant E-02/08 and the General
Research Funds Project No.~414911, established under the University
Grant Committee of the Hong Kong Special Administrative Region,
China. This work is also partially supported by the China 973 Program, Project No.~2012CB315904.}%
}

\markboth{}%
{Qing \MakeLowercase{\textit{et al.}}: Asynchronous Convolutional-Coded Physical-Layer Network Coding}


\maketitle

\begin{abstract}
This paper investigates the decoding process of asynchronous
convolutional-coded physical-layer network coding (PNC) systems.
Specifically, we put forth a layered decoding framework for
convolutional-coded PNC consisting of three layers: symbol realignment
layer, codeword realignment layer, and joint channel-decoding network coding
(Jt-CNC) decoding layer. Our framework can deal with phase asynchrony and
symbol arrival-time asynchrony between the signals simultaneously
transmitted by multiple sources. A salient feature of this framework is that
it can \lsc{handle} both fractional and integral symbol offsets; previously
proposed PNC decoding algorithms (e.g., XOR-CD and reduced-state Viterbi
algorithms) can only deal with fractional symbol offset. Moreover, the
Jt-CNC algorithm, based on belief propagation (BP), is BER-optimal for synchronous PNC and near optimal for asynchronous PNC.
Extending beyond convolutional codes, we further generalize the Jt-CNC decoding algorithm \lsc{for} all cyclic codes. Our simulation shows
that Jt-CNC outperforms the previously proposed XOR-CD algorithm and
reduced-state Viterbi algorithm by \unit[2]{dB} for synchronous PNC. \yq{For phase-asynchronous PNC, Jt-CNC is \unit[4]{dB} better than the other two algorithms}. \lsc{Importantly, for real wireless environment testing}, we \lsc{have also implemented} our decoding algorithm in a PNC system
built on the USRP software radio platform. Our experiment shows that the proposed Jt-CNC decoder
works well \lsc{in practice}.
\end{abstract}


%
\IEEEpeerreviewmaketitle

\section{Introduction}\label{sec:intro}

\IEEEPARstart{T}{his} paper investigates the use of convolutional codes in asynchronous
physical-layer network coding (PNC) systems to ensure reliable
communication. In particular, we focus on the decoding problem when
simultaneous signals from multiple transmitters arrive at a PNC receiver
with asynchronies between them.

PNC was first proposed in \cite{zhang2006hot} as a way to exploit
network coding \cite{ahlswede2000network,li2003linear} at the physical layer. In the simplest PNC setup, two users exchange
information via a relay in a two-way relay network (TWRN). The two users
transmit their messages simultaneously to the relay; the relay then maps the
overlapped signals to a network-coded message and broadcasts it to the two
users; and each of the two users recovers the message from the other user
based on the network-coded message and the knowledge of its own message. PNC
can potentially boost the throughput of TWRN by 100{\%} compared with a
traditional relay system \cite{zhang2006hot}.

Our paper focuses on PNC decoding as applied to TWRN. To ensure reliable
transmission, communication systems make use of channel coding to protect
the information from noise and fading. In channel-coded PNC, the goal of the
relay is to decode the simultaneously received signals not into the
individual messages of the two users, but into a network-coded message. This
process is referred to as the channel-decoding network coding (CNC) process
in \cite{zhang2009channel}.

In addition to the issue of channel coding, in practice, the signals from
the two users may be asynchronous in that there may be relative symbol
arrival-time asynchrony (symbol misalignment), phase asynchrony (phase offset), and other
asynchronies between the two signals received at the relay. These PNC
systems are referred to as asynchronous PNC (APNC) systems
\cite{lu2012asynchronous}.

Both \cite{zhang2009channel} and \cite{lu2012asynchronous} assume
the use of repeat accumulate (RA) codes. Our current paper, on the other
hand, focuses on the use of convolutional codes. A main motivation is that
convolutional codes are commonly adopted in many communications systems
(e.g., the channel code in IEEE $802.11$ is a convolutional code
\cite{lan2003part}). Convolutional codes have been well studied
and there are many good designs for the encoding/decoding of convolutional
codes in the conventional communication setting. Given this backdrop,
whether these designs are still applicable to PNC, and what additional
considerations and modifications are needed for PNC, are issues of utmost
interest. This paper is an attempt to address these issues.

Our main contributions are as follows:
\begin{itemize}
\item We put forth a layered decoding framework for asynchronous PNC system. The proposed decoding framework can deal with synchronous PNC as well as asynchronous PNC with relative phase offset and general symbol misalignment---by general symbol misalignment, we mean that the arrival times of the two users' signals at the relay are offset by $(\tau_{\mathrm{I}}+\tau_{\mathrm{F}})$ symbol durations, where $\tau_{\mathrm{I}}$  is an integral offset and $\tau_{\mathrm{F}}$ is a fractional offset smaller than one.
\item We design a joint channel-decoding network coding (Jt-CNC) decoder for convolutional-coded PNC. The Jt-CNC decoder, based on belief propagation (BP), is optimal in terms of bit error rate (BER) performance.
\item We implement the Jt-CNC decoder in a \lsc{real} PNC system built on USRP software radio platform. Our experiment shows that the Jt-CNC decoder works well under real wireless channel.
\item We propose an algorithm that can \lsc{handle} general symbol misalignment in cyclic-coded PNC, building on the insight obtained from our study of convolutional-coded \lsc{PNC; that} is, the algorithm is applicable to all cyclic codes, not just convolutional codes.
\end{itemize}

\lsc{The remainder of this paper} is organized as follows. Section \ref{sec:related}
overviews related work. Section \ref{sec:system} describes
the PNC system model. Section \ref{sec:synchronous} puts forth our
Jt-CNC framework, focusing on synchronous PNC. Section \ref{sec:asynchronous} extends the Jt-CNC framework to asynchronous PNC.
\lsc{We further show how the algorithmic framework is actually applicable to the general
cyclic-coded PNC}. Section \ref{sec:numerical} presents
simulations and experimental results. Section \ref{sec:conclusion}
concludes this work.

\section{Related Work}\label{sec:related}

\subsection{Synchronous PNC with Convolutional Codes}
The first implementation of TWRN based on the principle of PNC was recently
reported in \cite{lu2012implementation,lu2013real}. This system employs the
convolutional code defined in the 802.11 standard and adopts the OFDM
modulation to eliminate symbol misalignment \cite{rossetto2009design}. In \cite{lu2012implementation,lu2013real}, first the log-likelihood ratio (LLR) of the XORed channel-coded bits is computed; then
this soft information is fed to a conventional Viterbi decoder. We refer to
this decoding strategy as the soft XOR and channel decoding (XOR-CD) scheme
\cite{liew2013physical}. The experiment shows that the use of XOR-CD on the convolutional-coded
PNC system, thanks to its simplicity, \lsc{is feasible and practical}.

The acronym XOR-CD refers to a two-step process: first, prior to channel
decoding and without considering the correlations among the received symbols
due to the channel code, we apply symbol-by-symbol PNC mapping on the
received symbols to obtain estimates on the successive XORed bits; after
that, we perform channel decoding on the XORed bits to obtain the XORed
source bits. The performance of XOR-CD is suboptimal because the PNC mapping
in the first step loses information \cite{zhang2009channel}. Furthermore, only linear channel codes
can be correctly decoded in the second step. Jt-CNC, on the other hand,
performs channel decoding and network coding as an integrated process rather
than two disjoint steps. Jt-CNC can be ML (maximum likelihood) \lsc{optimal, depending on which variations of Jt-CNC we use and whether the underlying PNC system is synchronous or asynchronous.}

Within the class of Jt-CNC algorithms, for optimality, there are two
possible decoding targets: (i) ML XORed codeword; (ii) ML XORed bits. To
draw an analogy, for the conventional single-user point-to-point
communication, if convolutional codes are used, then the Viterbi algorithm
\cite{viterbi1967error} aims to obtain the ML codeword, while the BCJR
\cite{bahl1974optimal} aims to obtain ML bits. For PNC systems, the
aim is to obtain the network-coded codeword or the network-coded bits
instead.

A Jt-CNC algorithm for finding the XORed codeword was proposed in
\cite{to2010convolutional}. However, as will be discussed later,
finding the ML XORed codeword requires exhaustive search that could have
prohibitively high complexity. Therefore, the log-max approximation is
adopted in \cite{to2010convolutional} and the ML algorithm is simplified
to (approximated with) a full-state Viterbi algorithm. The term
``full-state'' comes from the fact that this algorithm combines the
trellises of both end nodes to make a virtual decoder. By searching the best
path on the combined trellis with the Viterbi algorithm, \cite{to2010convolutional} tries to decode
the ML pair of codewords of the two end nodes. To further reduce the
complexity, \cite{to2010convolutional} simplifies the full-state Viterbi
algorithm to a reduced-state Viterbi algorithm. Reference
\cite{to2010convolutional}, however, did not benchmark their approximate
algorithm with the optimal one. As we will show later, the algorithm
proposed by us in this paper can yield better performance than that in
\cite{to2010convolutional}.

In this paper, we aim to find the ML XORed bits within the codeword rather
than the overall ML XORed codeword. In Section \ref{sec:synchronous}
we show that our algorithm is ML XORed-bit optimal for synchronous PNC.
Finding ML XORed bits turns out to have much lower complexity than finding
the ML XORed codeword. This is quite different from the conventional
point-to-point communication system, in which the simple Viterbi algorithm
can be used to decode the ML codeword, and in which BCJR \lsc{(slightly more complex than the Viterbi algorithm)} can be used to decode the ML bits.

\subsection{Asynchronous PNC with Convolutional Codes}
In asynchronous PNC systems, the signals from the two end nodes may arrive
at the relay with symbol misalignment and relative phase offset \cite{lu2012asynchronous}. To our
best knowledge, there was no Jt-CNC decoder for convolutional codes that can
deal with integral-plus-fractional symbol misalignment. In \cite{wang2009channel}, a convolutional
decoding scheme with an XOR-CD algorithm was proposed to deal with integral
symbol misalignment. As pointed out in \cite{wang2009channel}, symbol misalignment entangles
the channel-coded bits of the trellises of the two encoders in a way that
ordinary Viterbi decoding, based on just one of the trellises, is not
applicable. Therefore the XOR-CD algorithm for synchronous PNC cannot be
applied anymore in the presence of integral symbol misalignment. Their
solution is to rearrange the transmit order of the channel-coded bits into
blocks, and pad $D_{\max}$ zeros between adjacent blocks. The zero padding
acts as a guard interval between blocks that avoids the entanglement of
channel-coded bits and facilitates Viterbi decoding. However, this scheme
can only deal with integral symbol misalignment of at most $D_{\mathrm{max}}$
symbols. In addition, it incurs a code-rate loss factor of $(1-D_{\mathrm{max}}/L)$ due to the zero padding between blocks.

\subsection{Asynchronous PNC with Other Channel Codes}
\label{subsec:asynchronous}
The use of LDPC codes in asynchronous PNC systems have previously been
considered. In \cite{lu2012asynchronous}, the authors designed a Jt-CNC
decoder for the RA code that can deal with fractional
symbol misalignment (i.e., symbol misalignment \lsc{that} is less than one symbol
duration) and phase offset. Our decoding framework adopts the over-sampling
technique proposed in \cite{lu2012asynchronous} to address fractional
symbol misalignment.

To deal with asynchrony in PNC, our decoding framework consists of three
layers: symbol-realignment layer, codeword-realignment layer, and joint
channel-decoding network coding (Jt-CNC) layer. The \lsc{first two}
layers, symbol realignment and codeword realignment, counter fractional and
integral symbol misalignments, respectively; the \lsc{third} layer, Jt-CNC,
decodes the ML XORed bits. Other decoding schemes (e.g., XOR-CD, full-state
Viterbi) can also be used in the third layer of the framework.
We further show that \lsc{our decoding framework is not only applicable when convolutional codes are adopted, it is also applicable when general cyclic codes are used. Besides convolutional codes, an important class of cyclic codes is the cyclic LDPC.}

The Jt-CNC decoder proposed in \cite{lu2012asynchronous} was extended by
\cite{wu2011joint} to deal with general asynchrony using cyclic
LDPC. However, the proposed decoder in \cite{wu2011joint} discards the non-overlapped part
of the received signal, losing useful information that can potentially
enhance performance. \lsc{Therefore, for the decoder in} \cite{wu2011joint}, the larger the symbol misalignment, the worse the performance. \lsc{By contrast}, our framework
makes full use of the non-overlapped \lsc{portion of the signal} so that the
\lsc{larger} symbol misalignment can enhance performance.

\section{System Model}\label{sec:system}

\begin{figure}[htb]
\centering
\includegraphics[width=12cm]{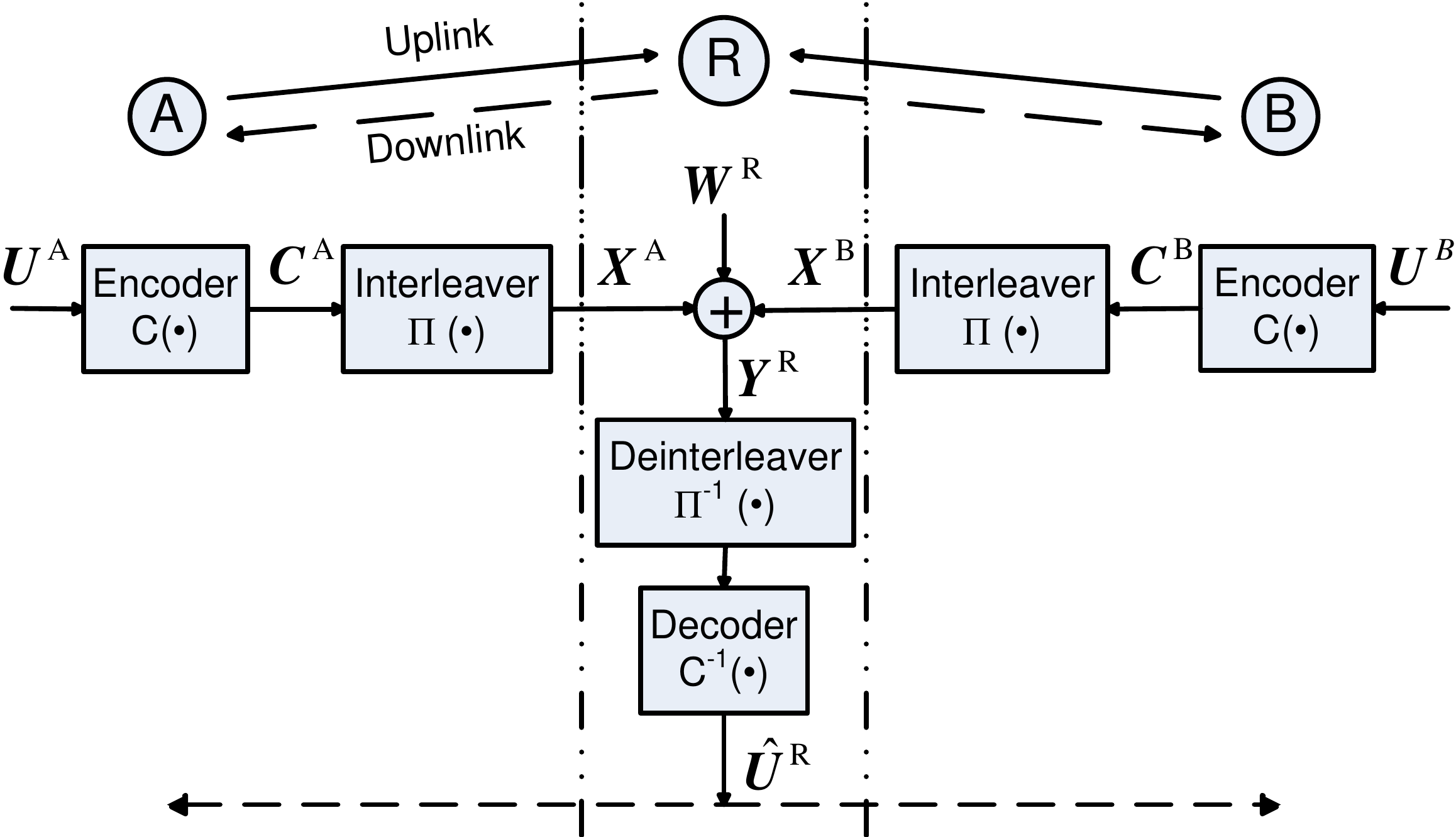}\\
\caption{System model of two-way relay network operated with physical-layer network coding.}\label{fig1}
\end{figure}

We consider the application of PNC in a two-way relay network (TWRN) as
shown in Fig. \ref{fig1} In this model, nodes A and B exchange information with the
help of relay node R. We assume that all nodes are half-duplex and there is
no direct link between A and B.

With PNC, nodes A and B exchange one packet with each other in two time
slots. The first time slot corresponds to an \emph{uplink phase}, in which node A
and node B transmit their channel-coded packets simultaneously to relay R.
The relay R then constructs a network-coded packet based on the
simultaneously received signals from A and B. This operation is referred to
as the channel decoding network coding (CNC) process \cite{liew2013physical}, because the
received signals are decoded into a network-coded message rather than the
individual messages from A and B. The second time slot corresponds to a
\emph{downlink phase}, in which relay R channel-codes the network-coded message and
broadcasts it to both A and B. Upon receiving the network-coded packet, A
(B) then attempts to recover the original packet transmitted by B (A) in the
uplink phase using self-information \cite{zhang2006hot}. This paper
focuses on the design of the CNC algorithm in the uplink phase; the issue in
the downlink phase is similar to that in conventional point-to-point
transmission and does not require special treatment \cite{liew2013physical}.

As shown in Fig. \ref{fig1}, in the uplink phase, the
source packets of nodes A and B each goes through a convolutional encoder,
an interleaver, and a modulator. We adopt tail biting convolutional code\footnote{The use of other kinds of convolutional codes (e.g., zero-tailing and recursive) is discussed in the Appendix.}
\cite{ma1986tail} and block interleaver throughout this paper.
We denote the source packets of node A and node B by two $K$-bit binary
sequences:
\begin{equation}
\label{eq1}
U^i=\left( {u_1^i ,u_2^i ,\cdots ,u_K^i } \right),\  i \in \left\{
{\mathrm{A,B}} \right\}
\end{equation}
where $u_k^i$ is the input bit of end nodes $i$'s source packet at time $k$. The
source packets are encoded into two $M$-bit channel-coded binary sequences. We
assume nodes A and B use the same convolutional code with code rate $r=1/R$
where $R$ is an integer. In the following presentation we choose $r=1/3$,
thus $M=3K$. The two channel-coded packets are
\begin{eqnarray}
\label{eq2}
 C^i&=&\left( {c_1^i ,c_2^i ,\cdots ,c_M^i } \right) \nonumber \\
 &=&\left( {\bar {c}_1^i ,\bar {c}_2^i ,\cdots ,\bar {c}_K^i } \right) \nonumber \\
 &=&\left( {c_{1,1}^i ,c_{1,2}^i ,c_{1,3}^i,\;c_{2,1}^i ,c_{2,2}^i ,\cdots
,c_{K,1}^i ,c_{K,2}^i ,c_{K,3}^i } \right),\  i\in \left\{
{\mathrm{A,B}} \right\}
\end{eqnarray}
where $c_{k,j}^i$ is the $j$th channel-coded bit of end nodes $i$'s
channel-coded packet at time $k$; the 3-bit tuple $\bar{c}_k^i =(c_{k,1}^i
,c_{k,2}^i ,c_{k,3}^i)$ is the output of the convolutional encoder of node
$i$ at time $k$. Then, \lsc{$C^\mathrm{A}$ and $C^\mathrm{B}$} are fed into block interleavers that realize the same
permutation to produce
\begin{equation}
\label{eq3}
\tilde{C}^i=\left( {c_{1,1}^i ,c_{2,1}^i ,\cdots ,c_{K,1}^i ,\;c_{1,2}^i
,\cdots ,c_{K,2}^i ,\;c_{1,3}^i ,\cdots ,c_{K,3}^i } \right),\  i\in
\left\{ \mathrm{A},\mathrm{B} \right\}.
\end{equation}

Note that the permutation groups the $j$th coded bits of all times together into a block. There are altogether three blocks. Finally, $\tilde{C}^i$ are modulated to produce the two sequences of $N$ complex
symbols:
\begin{equation}
\label{eq4}
X^i=\left( {x_1^i ,x_2^i ,\cdots ,x_N^i } \right),\  i \in \left\{
\mathrm{A,B} \right\}.
\end{equation}

Throughout this paper, we focus on BPSK and QPSK modulations; our framework
can be easily extended to higher order constellations \cite{namboodiri2013physical,yang2010modified,koike2009optimized}. For BPSK $N{=}3K$ and
$x_n^i \in \left\{1,-1\right\}$. For QPSK $N{=}3K/2$ and $x_n^i {\in} 1/{\sqrt 2 }\left\{ {1+j,-1+j,1-j,-1-j} \right\}$.
The complex symbol sequences \yq{$X^\mathrm{A}$ and $X^\mathrm{B}$} are shaped using a pulse shaping function $p(t)$ with symbol
duration $T$ and transmitted. Without loss of generality, we assume $p(t)$ is
the rectangular pulse throughout this paper.

Let us denote the channel coefficients of the channels from node A and node
B to relay R by $h^\mathrm{A}$ and $h^\mathrm{B}$, respectively. Both
$h^\mathrm{A}$ and $h^\mathrm{B}$ are complex numbers, whose phase difference
$\phi=\angle(h^\mathrm{B}/h^\mathrm{A})$ is the relative phase offset
between node A and node B. We assume that the channel state information (CSI)
$h^\mathrm{A}$ and $h^\mathrm{B}$ can be estimated at the relay R using
orthogonal preambles \cite{lu2012implementation}.

The received complex baseband signal at the relay is
\begin{equation}
\label{eq5}
y^\mathrm{R}(t)=\sum\limits_{n=1}^N {\left\{ {h^\mathrm{A}x_n^\mathrm{A} p\left(
{t-nT} \right)+h^\mathrm{B}x_n^\mathrm{B} p\left( {t-nT-\tau T} \right)}
\right\}} +w^\mathrm{R}(t)
\end{equation}
where $\tau T$ is the symbol misalignment (i.e., the arrival time of the
signal of B lags the arrival time of the signal of A by $\tau T$), and
$w^\mathrm{R}(t)$ is the noise, assumed to be circularly complex with variance
$\sigma^2$. We assume the symbol misalignment to consist of two parts: an
integral part $\tau_\mathrm{I} \in \mathbf{N}^{+}$; and a fractional part $\tau
_\mathrm{F} \in [0,1)$ so that $\tau =\tau _\mathrm{I} +\tau _\mathrm{F}$.

\section{Synchronous Convolutional-Coded PNC}\label{sec:synchronous}

This section focuses on synchronous convolutional-coded PNC, where the
signals of node A and node B are symbol-aligned ($\tau =0)$. We first derive
the \yq{XOR packet-optimal} Jt-CNC algorithm that aims at finding the ML XORed
source packet. We show that the XOR packet-optimal algorithm has
prohibitively high complexity. Then we introduce our \yq{XOR bit-optimal} Jt-CNC
algorithm \lsc{for} finding the ML XORed bits, which has much lower complexity.

\subsection{XOR Packet-Optimal Decoding of Synchronous Convolutional-Coded PNC}
In the case of synchronous convolutional-coded PNC, the received baseband
signal at relay R is obtained by setting symbol misalignment $\tau$ to zero
in (\ref{eq5}):
\begin{equation}
\label{eq6}
y^\mathrm{R}(t)=\sum_{n=1}^N \left\{ h^\mathrm{A}x_n^\mathrm{A} p(t-nT)+h^\mathrm{B}x_n^\mathrm{B}p(t-nT) \right\}
+w^\mathrm{R}(t).
\end{equation}

After matched filtering \cite{lu2012asynchronous}, the received baseband samples at relay R are
\begin{equation}
\label{eq7}
Y^\mathrm{R}=\left(y_1^\mathrm{R}, y_2^\mathrm{R}, \cdots, y_N^\mathrm{R}\right)
\end{equation}
where
\begin{equation}
\label{eq8}
y_n^\mathrm{R} = h^\mathrm{A}x_n^\mathrm{A} + h^\mathrm{B}x_n^\mathrm{B} +w_n^\mathrm{R}.
\end{equation}

The ML XORed source packet $\hat{U}^\mathrm{R}=(\hat{u}_1^\mathrm{R} ,\hat{u}_2^\mathrm{R}, \cdots, \hat{u}_K^\mathrm{R})$
(i.e., ML XOR of the source packets of node A and node B) is given by
\begin{equation}
\label{eq9}
\hat{U}^\mathrm{R}=\arg \max_{\scriptscriptstyle U^\mathrm{R}}
\sum_{\scriptscriptstyle U^\mathrm{A},U^\mathrm{B}:U^\mathrm{A}\oplus U^\mathrm{B}=U^\mathrm{R}}
{\exp \left( {-\mathcal{M}\left( {X^\mathrm{A},X^\mathrm{B}} \right)} \right)}
\end{equation}
where $\oplus$ denotes the binary bit-wise XOR operator; $X^\mathrm{A}$ and
$X^\mathrm{B}$ are the convolutional-encoded and modulated baseband signal
of $U^\mathrm{A}$ and $U^\mathrm{B}$, respectively; \lsc{and $\mathcal{M}(X^\mathrm{A},X^\mathrm{B})$ is the distance metric defined as follows:}
\begin{eqnarray}
\label{eq10}
\mathcal{M}\left(X^\mathrm{A},X^\mathrm{B}\right)&=&\sum_{n=1}^N
\frac{\left| y_n^\mathrm{R} -h^\mathrm{A}x_n^\mathrm{A} -h^\mathrm{B}x_n^\mathrm{B} \right|^2}{2\sigma^2} \nonumber\\
 &=&\frac{\left\| {Y^\mathrm{R}-h^\mathrm{A}X^\mathrm{A}-h^\mathrm{B}X^\mathrm{B}}
\right\|_2^2 }{2\sigma ^2}.
\end{eqnarray}

For source packets $U^{\mathrm{A}}$ and $U^{\mathrm{B}}$ of length $K$,
the functional mapping from $U^{\mathrm{A}}$ and $U^{\mathrm{B}}$
to the XORed source packet $U^{\mathrm{R}}$ can be expressed as

\begin{equation}\label{eq:xor_map1}
f_{\mathrm{packet}} :\{0,1\}^K\times \{0,1\}^K \to \{0,1\}^K.
\end{equation}

The mapping in (\ref{eq:xor_map1}) is a $2^K$-to-$1$ mapping; that is, there are $2^K$ possible $(U^{\mathrm{A}},U^{\mathrm{B}})$ that can produce a particular $U^{\mathrm{R}}$. This is where the complexity lies in (\ref{eq9}). For each $X^{\mathrm{R}}$, the baseband channel-coded signal corresponding to $U^{\mathrm{R}}$, we need to examine $2^K$ possible combinations of $X^{\mathrm{A}}$ and $X^{\mathrm{B}}$. The Viterbi algorithm is a shortest-path algorithm that computes a path in the trellis of $(U^{\mathrm{A}},U^{\mathrm{B}})$. Meanwhile, each $U^{\mathrm{R}}$ is associated with $2^K$ paths in the trellis. There is no known exact computation method for (\ref{eq9}) except to exhaustively sum over the   possible combinations of $(U^{\mathrm{A}},U^{\mathrm{B}})$ for each $U^{\mathrm{R}}$.

\yq{
For each possible $(U^{\mathrm{A}},U^{\mathrm{B}})$, we need to sum over $N$ terms in (\ref{eq10}) to compute $\mathcal{M}(X^\mathrm{A},X^\mathrm{B})$. For a code-rate $r$ code and M-QAM modulation, $N = K/[r{\log_2}(M)]$. Computing each term in (\ref{eq10}) takes two complex operations, and the summation takes $(N - 1)$ operations. Hence the complexity of one combination of $(U^{\mathrm{A}},U^{\mathrm{B}})$ is $(3K/[r{\log _2}(M)] - 1)$. Moreover, to find the maximum of (\ref{eq9}), $({2^K} - 1)$ comparisons are needed. Given that there are $2^K$ possible $U^{\mathrm{R}}$, from which we want to find the optimal $\hat{U}^\mathrm{R}$, the overall complexity is therefore ${2^{2K}}(3K/[r{\log _2}(M)] - 1) + {2^K} - 1$. In Big-O notation, the complexity is $O(K{2^{2K}})$.
}

This is a big contrast with the regular point-to-point communication system, in which the Viterbi algorithm used to find the ML codeword is of polynomial complexity only. For PNC systems, the complexity of XOR packet-optimal decoding algorithm is exponential with $K$, the length of source packet.

\subsection{XOR Bit-Optimal Decoding of Synchronous Convolutional-Coded PNC}
\label{subsec:mylabel1}

To reduce complexity, we consider an XOR bit-optimal Jt-CNC decoder based on the framework of Belief Propagation (BP) algorithms. The proposed decoder aims to find the ML XORed source bit rather than the ML XORed source packet. We give two important results: (i) the proposed Jt-CNC decoder is optimal in terms of BER performance; and (ii) the complexity is linear in packet length $K$.

Unlike finding ML XORed packets, for which the Viterbi algorithm is of little use, the BP (BCJR) algorithm can find the ML XORed source bit readily without incurring exponential growth in complexity. We first explain the reason before describing the BP algorithm in detail.

The $k$th ML XORed source bits $\hat u_k^{\mathrm{R}},k = 1,2, \ldots K$ is given by
\begin{equation}\label{eq:mlxorbit}
    \hat u_k^{\mathrm{R}} = \arg \max_{u_k^{\mathrm{R}}} \sum_{{{\bar u}_k}:u_k^{\mathrm{A}} \oplus u_k^{\mathrm{B}} = u_k^{\mathrm{R}}} \Pr \left( {{\bar u}_k}| {Y^{\mathrm{R}}} \right)
\end{equation}
where $\Pr (\bar{u}_k|Y^{\mathrm{R}} )$ can be calculated using the BP algorithm. Fortunately, finding the ML XORed bits in PNC systems has much lower complexity, because the functional mapping from $(u_k^\mathrm{A},u_k^\mathrm{B})$ to $u_k^{\mathrm{R}}$ can be expressed as
\begin{equation}\label{eq:xorbitmap}
    f_{{\mathrm{bit}}}:\{ 0,1\}  \times \{ 0,1\}  \to \{ 0,1\}.
\end{equation}

The mapping in (\ref{eq:xorbitmap}) is a $2$-to-$1$ mapping; hence for each possible XOR bit we need to examine only two pairs of source bits. Importantly, the BP algorithm can compute $\Pr(u_k^\mathrm{A},u_k^\mathrm{B}| Y^{\mathrm{R}})$ easily, from which $\Pr(u_k^\mathrm{A} \oplus u_k^\mathrm{B}| Y^{\mathrm{R}})$ can readily be obtained through the $2$-to-$1$ mapping. Because finding a single ML XORed source bit in (\ref{eq:mlxorbit}) takes two summations and one comparison, so finding all the ML XORed source bits takes $3K$ operations beyond the operations by the BP algorithm that computes $\Pr(u_k^\mathrm{A},u_k^\mathrm{B}|Y^{\mathrm{R}}), k = 1,2, \ldots K$.

\yq{
As will be elaborated later in this section, the BP algorithm has three steps. The initialization takes $9K/[rlog_{2}(M)]$ operations in (\ref{eq14}). The forward/backword recursions take  $6 {\cdot} {2^{2/r}}K{S^2}$ operations where $S$ is the number of decoder's states. The termination takes $4{\cdot} {2^{2/r}} K{S^2}$ operations. Therefore finding the ML XORed source bits of length-$K$ packets has an overall complexity of $9K/[r log_{2}(M)] + 6 {\cdot} {2^{2/r}}K{S^2} + 4{\cdot} {2^{2/r}}K{S^2} + 3K$. In Big-O notation, the complexity of source bit-optimal decoding algorithm is $O(K)$. We next elaborate the bit-optimal Jt-CNC algorithm.}

BP is a framework for generating inference-making algorithms for graphical
models, in which there are two kinds of nodes: variable nodes and factor
nodes. Each variable node represents a variable, such as the state variable
of the convolutional encoder; each factor node indicates the relationship
among all variable nodes connected to it. For example the state transition
function of a convolutional encoder is represented by a factor node.
The goal of BP is to compute the marginal probability distributions $\Pr(u_k^\mathrm{A},u_k^\mathrm{B}| Y^{\mathrm{R}})$ for all $k$. This goal is achieved by means of a
sum-product message-passing algorithm \cite{pearl1988probabilistic}.

\begin{figure}[htb]
\centering
  \includegraphics[width=12cm]{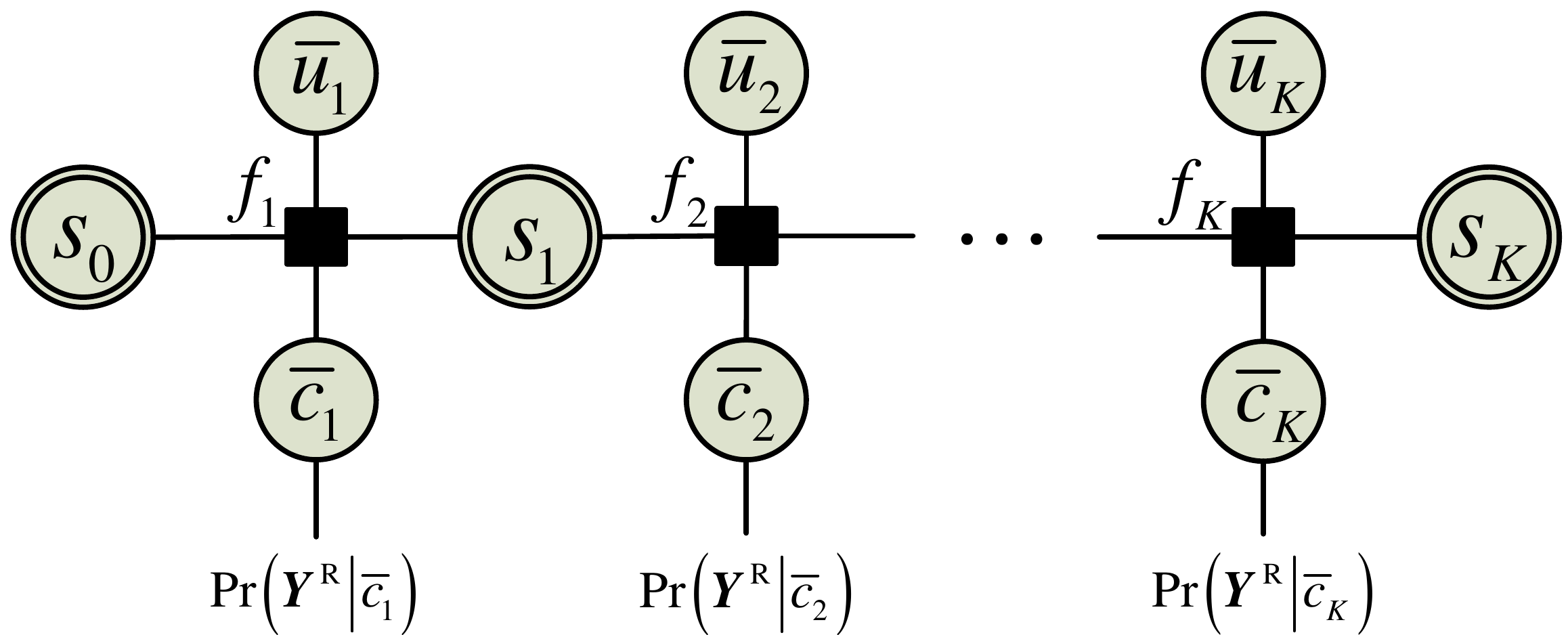}\\
  \caption{Tanner graph of the Jt-CNC decoder on which the BP algorithm operates: $s_k$ is the state variable, $\bar{u}_k$ is the source bit, $\bar{c}_k$ is the channel-coded bit; $f_i$ is the factor node that represents the state transition function of the Jt-CNC decoder.}\label{fig2}
\end{figure}

Fig. \ref{fig2} shows the Tanner graph of our bit-optimal Jt-CNC decoder. Unlike the
conventional point-to-point convolutional decoder for single-user systems
with only one transmitter, the Jt-CNC decoder combines the states and the
trellis of both transmitters A and B. In Fig. \ref{fig2}, vectors $S{=}( {s_0 ,s_1
,\cdots ,s_K })$ represents the state variables, where state $s_k$
combines the state of both end nodes' states; vector $U=( \bar{u}_1, \bar{u}_2, \cdots, \bar{u}_K)$, where $\bar{u}_k=(u_k^\mathrm{A}, u_k^\mathrm{B})$, represents the ``virtual'' source packet consisting of the duple of the two source packets from nodes A and B;
similarly, vector $C=(\bar{c}_1, \bar{c}_2, \cdots, \bar{c}_K)$, where $\bar{c}_k=(\bar{c}_k^\mathrm{A}, \bar{c}_k^\mathrm{B})$ (as defined in (\ref{eq2}) $\bar{c}_k^i $ denotes the
group of channel-coded bits of node $i$ at time $k$), represents the ``virtual''
channel-coded packet, assuming that both nodes A and B use the same channel
code. The behavior of the decoder is defined by the functions of the factors
node $f_k(s_{k-1}, \bar{u}_k, \bar{c}_k, s_k)$ that
represents the state transition rule of the trellis.

The goal of the Jt-CNC decoder is to find the maximum likelihood XOR bit $u_k^\mathrm{R}$ through the \emph{a posteriori probability} (APP) $\Pr (\bar{u}_k \vert Y^\mathrm{R})$ by
\begin{equation}
\label{eq11}
\Pr\left(u_k^\mathrm{R} \left|
Y^\mathrm{R}\right. \right)=\max_{u_k^\mathrm{R}}
\sum_{\bar{u}_k:u_k^\mathrm{A} \oplus u_k^\mathrm{B}=u_k^\mathrm{R}}
{\Pr \left( {\bar{u}_k \left| {Y^\mathrm{R}} \right.} \right)}
\end{equation}
where $\Pr (\bar{u}_k \vert Y^\mathrm{R})$ can be computed exactly by the
sum-product message-passing algorithm thanks to the tree structure of the
Tanner graph associated with convolutional nodes
\cite{kschischang2001factor}. The sum-product algorithm, when applied to
decode convolutional codes, is the well-known BCJR algorithm
\cite{bahl1974optimal}. The difference in our situation here is that
instead of the source bit from one source, we are decoding for the bit duple
$\bar{u}_k =(u_k^\mathrm{A} ,u_k^\mathrm{B})$ from the two sources.

\begin{figure}[htb]
\centering
  \includegraphics[width=12cm]{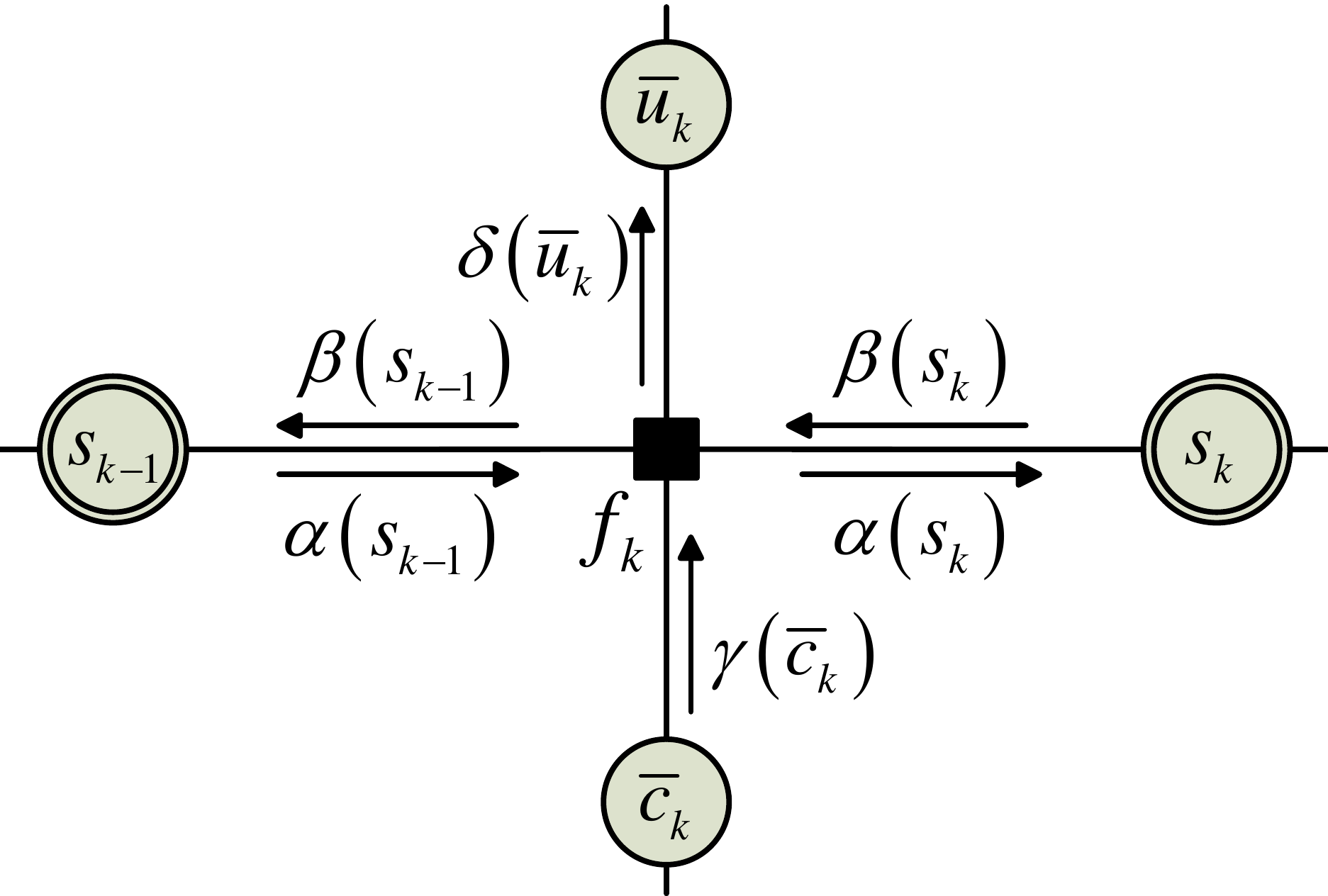}\\
  \caption{The messages being passed around a factor node during the operation of the sum-product algorithm.}\label{fig3}
\end{figure}

We now explain the sum-product algorithm in detail.
Fig. \ref{fig3} depicts the messages being passed around
a factor node within the overall Tanner graph of
Fig. \ref{fig2}. We follow the notation of the original
paper on the BCJR algorithm \cite{bahl1974optimal}. In the forward
direction, the message from $s_{k-1}$ to $f_k$ is denoted by $\alpha
(s_{k-1})$, and the message from $f_k$ to $s_k$ is denoted by $\alpha
(s_k )$. In the backward direction, the message from $s_k$ to $f_k$ is
denoted by $\beta(s_k )$, and the message from $f_k$ to $s_{k-1}$ is
denoted by $\beta(s_{k-1})$. Additionally, $\gamma(\bar{c}_k)$ denotes
the message from $\bar{c}_k$ to $f_k$, and $\delta(\bar{u}_k)$ denotes
the message from $f_k$ to $\bar{u}_k$. Note that $\delta(\bar{u}_k)$ is the APP $\Pr (\bar{u}_k \vert Y^\mathrm{R})$ and the goal
here is to compute it.

Since the Tanner graph of the Jt-CNC decoder is cycle-free, the operation of
the sum-product algorithm consists of two natural recursions according to
the direction of message flow in the graph: a \emph{forward recursion} to compute
$\alpha(s_k)$ as a function of $\alpha(s_{k-1})$ and $\gamma(\bar{c}_k
)$; a \emph{backward recursion} to compute $\beta(s_{k-1})$ as a function of
$\beta(s_k )$ and $\gamma(\bar {c}_k)$.

The calculation of $\Pr (\bar{u}_k \vert Y^\mathrm{R})$ can be divided into
three steps: initialization, forward/backward recursion, and termination. We
present these three steps in detail below.

\emph{Initialization} As usual in a cycle-free Tanner graph, the sum-product algorithm begins at
the leaf nodes. Since tail biting convolutional code is used, the initial
and terminal states of end node's convolutional encoders are the same, and
they are decided by the random input message. Therefore the initial and
terminal states are uniformly distributed among all possible states, so the
message $\alpha(s_0)$ and $\beta(s_K)$ are initialized as
\begin{subequations}
\begin{align}
\label{eq12}\alpha(s_0)&=\frac{1}{N_S},\  \forall s_0\\
\intertext{and}
\label{eq13}\beta (s_K )&=\frac{1}{N_S },\  \forall s_K
\end{align}
\end{subequations}
where $N_S $ is the number of states per stage.

The message $\gamma(\bar{c}_k)$ is the likelihood function of $\bar{c}_k$ based on the evidence $Y^\mathrm{R}$. For example, if the code rate is $1/3$
and the BPSK modulation is used, then $\bar{c}_k =(c_{k,1}^\mathrm{A}
,c_{k,2}^\mathrm{A} ,c_{k,3}^\mathrm{A} ,c_{k,1}^\mathrm{B} ,c_{k,2}^\mathrm{B}
,c_{k,3}^\mathrm{B})$. These channel-coded bits $\bar{c}_k$ are
mapped to BPSK modulated symbols $(x_{3k-2}^\mathrm{A}
,x_{3k-1}^\mathrm{A} ,x_{3k}^\mathrm{A})$ and $(x_{3k-2}^\mathrm{B} ,x_{3k-1}^\mathrm{B} ,x_{3k}^\mathrm{B})$ at node A
and node B, respectively. Given the overlapped signal $Y^\mathrm{R}$ at the
relay node, the likelihood of $\bar{c}_k$ is calculated by
\begin{equation}
\label{eq14}
\gamma(\bar{c}_k)=\Pr \left(Y^\mathrm{R}\left| \bar{c}_k \right.
\right)=\prod_{j=1}^3\frac{1}{\sqrt{2\pi\sigma^2}}\exp \left\{
{-\frac{\left| {y_{3(k-1)+j}^\mathrm{R} -h^\mathrm{A}x_{3(k-1)+j}^\mathrm{A}
-h^\mathrm{B}x_{3(k-1)+j}^\mathrm{B} } \right|^2}{2\sigma ^2}} \right\}.
\end{equation}

\emph{Forward/backward recursion} After initializing the messages from leaf nodes, we can compute the message
$\alpha(s_k)$ and $\beta(s_k)$ recursively by following the message
update rule below \cite{kschischang2001factor}:
\begin{subequations}
\begin{align}
\label{eq15}
\alpha\left(s_k \right)&=\sum_{s_{k-1} ,\bar{u}_k ,\bar{c}_k}
{f_k \left( {s_{k-1},\bar{u}_k ,\bar{c}_k ,s_k} \right)} \alpha\left(
{s_{k-1} } \right)\gamma \left( {\bar {c}_k } \right)\\
\label{eq16}
\beta \left( {s_{k-1} } \right)&=\sum\limits_{s_k ,\bar {u}_k ,\bar {c}_k }
{f_k \left( {s_{k-1} ,\bar {u}_k ,\bar {c}_k ,s_k } \right)} \beta \left(
{s_k } \right)\gamma \left( {\bar {c}_k } \right).
\end{align}
\end{subequations}

\emph{Termination} In the final step, the algorithm terminates with the computation of $\delta(\bar{u}_k)$, which gives the APP of the source bit $\bar{u}_k$.
\begin{equation}
\label{eq17}
\delta\left(\bar{u}_k \right)=\sum_{s_{k-1},s_k,\bar{c}_k}
{f_k \left( {s_{k-1} ,\bar {u}_k ,\bar {c}_k ,s_k } \right)} \alpha \left(
{s_{k-1} } \right)\gamma \left( {\bar {c}_k } \right)\beta \left( {s_k }
\right).
\end{equation}

The summation in (\ref{eq17}) is over different trellis transitions $e=( {s_{k-1}
,\bar{u}_k, \bar{c}_k, s_k})$ with fixed $\bar{u}_k$, such that
$f_k(e)=1$ if $e$ is a valid transition, and $f_k(e)=0$ otherwise. For example, if input $\bar{u}_k$ causes a state
transition from $s_{k-1}$ to $s_k$ and the output is $\bar{c}_k$, then
$f_k(e)=1$; on the other hand, if input $\bar{u}_k$ causes a
state transition from $s_{k-1}$ to a state not equal to $s_k$ or the
output is not $\bar{c}_k$, then $f_k(e)=0$.

\section{Asynchronous Convolutional-Coded PNC}\label{sec:asynchronous}

\begin{figure}[htb]
\centering
  \includegraphics[width=6cm]{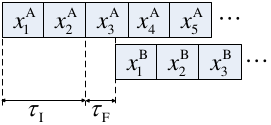}\\
  \caption{Symbol misalignment in PNC: a general symbol misalignment consists of an integral part $\tau_{\mathrm{I}}=2$ and a fractional part $\tau _{\mathrm{F}}=0.7$.}\label{fig4}
\end{figure}

In this section, we present our three-layer decoding framework for
asynchronous convolutional-coded PNC. The asynchrony causes unique
challenges that the synchronous decoder in Section \ref{sec:synchronous} cannot handle. As
shown in Fig. \ref{fig4}, when the signals of nodes A
and B arrive at the relay at different times, their symbols can be
misaligned. The symbol misalignment consists of two parts: an integral part
$\tau_{\mathrm{I}}$ and a fractional part $\tau _{\mathrm{F}}$. These two components impose different challenges: the fractional symbol misalignment causes overlaps of
adjacent symbols so that the symbol-boundary preserving sampling as
expressed in (\ref{eq8}) is no more valid; the integral symbol misalignment
entangles the channel-coded bits of nodes A and B in such a way that the
decoding scheme as proposed in Section IV cannot be
applied anymore.

\begin{figure}[htb]
\centering
  \includegraphics[width=12cm]{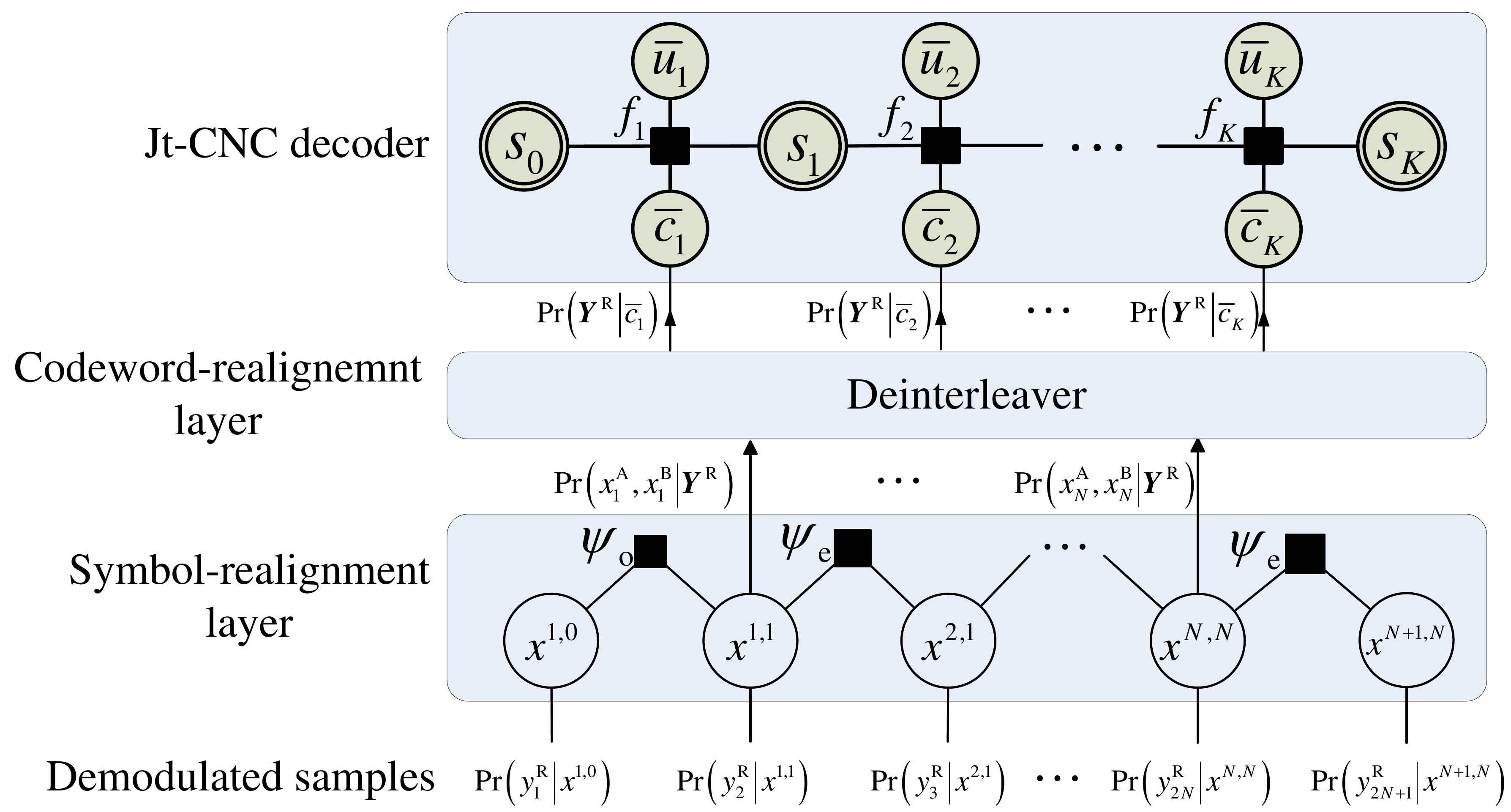}\\
  \caption{Decoding framework for asynchronous convolutional-coded PNC systems. This framework can deal with an integral-plus-fractional symbol misalignment and is BER-optimal.}\label{fig5}
\end{figure}

To address these challenges, together with the Jt-CNC decoder, we add two
layers to construct an integrated framework illustrated in
Fig. \ref{fig5}. First, to address the fractional symbol
misalignment, the symbol-realignment layer uses a BP algorithm at the relay
to ``realign'' the soft information of the symbols. Second, the
codeword-realignment layer uses an interleaver/deinterleaver set-up to
accommodate the integral symbol misalignment. As a result, the three-layer
decoding framework can deal with the integral-plus-fractional symbol
misalignment.

Furthermore, building on the insight obtained from our study of
convolutional-coded PNC, we propose an algorithm that can deal with general
symbol misalignment with cyclic codes. That is, our decoding framework can
incorporate not just convolutional codes, but all cyclic codes to address
the challenges in asynchronous PNC.

\subsection{Symbol-Realignment Layer: Addressing Fractional Symbol Misalignment }
\label{subsec:symbol}
For simplicity, as in \cite{lu2012asynchronous}, we assume the use of
rectangular pulse to carry the modulated signal, and the use of the doubling
sampling technique to obtain two samples per symbol period. Let us first
ignore the integral part of symbol misalignment and only consider the
fractional part (i.e., $\tau<1$). Furthermore, let us assume $|h_\mathrm{A}|=|h_\mathrm{B}|=\sqrt P$ where $P$ is the
transmission power of end nodes.

With double sampling on the received signal $y^\mathrm{R}(t)$, the total
number of samples obtained per frame is $2N+1$, where $N$ is the number of
symbols per frame (for both users A and B). The relay uses the $2N+1$ samples to compute the soft information $\Pr(x_n^\mathrm{A},x_n^\mathrm{B}
\vert Y^\mathrm{R})$, where instead of the expression in (\ref{eq7}),
$Y^\mathrm{R}=(y_1^\mathrm{R} ,y_2^\mathrm{R} ,\cdots ,y_{2N+1}^\mathrm{R})$ consists of the $2N+1$ samples. Thus, as far as the soft
information is concerned, \emph{the fractional symbol misalignment is removed and the symbols are realigned}. We emphasize that this realignment of soft
information is a key step. Once that is done, the channel decoding algorithm
for synchronous PNC as proposed in Section IV can
be applied.

We can write the samples obtained at the relay R as follows (after
normalization):
\begin{subequations}
\begin{align}
\label{eq181}
y_{2n-1}^\mathrm{R} &=x_n^\mathrm{A} +x_{n-1}^\mathrm{B} e^{j\varphi
}+w_{2n-1}^\mathrm{R} \\
\label{eq182}
y_{2n}^\mathrm{R} &=x_n^\mathrm{A} +x_n^\mathrm{B} e^{j\varphi }+w_{2n}^\mathrm{R}
\end{align}
\end{subequations}
where $n{=}1,2,\ldots,N$, $x_0^\mathrm{B} {=}0$ and $y_{2N+1}^\mathrm{R}
{=}x_N^\mathrm{B} e^{j\varphi }+w_{2N+1}^\mathrm{R} $. The terms
$w_{2n-1}^\mathrm{R} $ and $w_{2n}^\mathrm
{R}$ are zero-mean complex Gaussian
noise with variances ${\sigma ^2}/{\tau P}$ and ${\sigma ^2}/{(1-\tau )P}$ per dimension, respectively.

\begin{figure}
\centering
  \includegraphics[width=12cm]{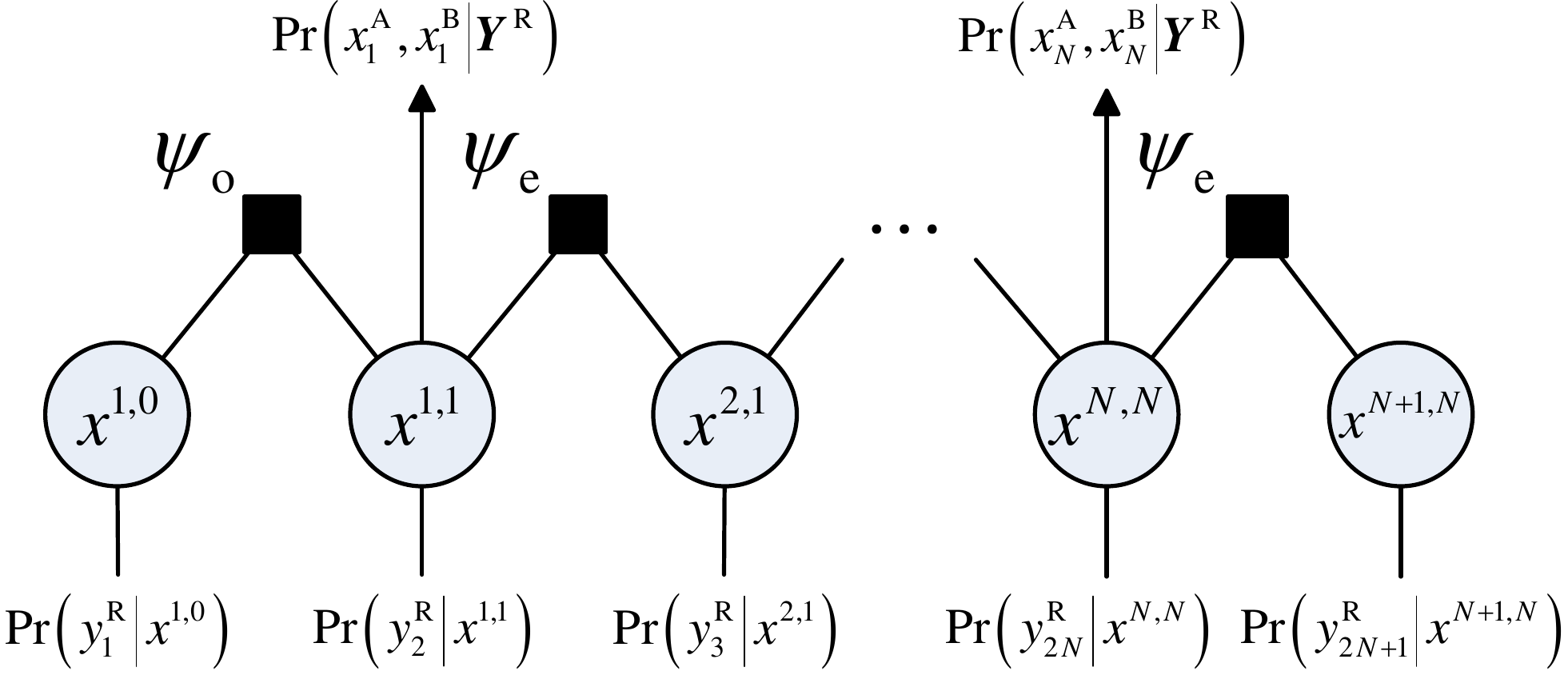}\\
  \caption{Tanner graph of the symbol-realignment layer. $x^{i,j}$, denotes the joint symbol $(x_i^\mathrm{A} ,x_j^\mathrm{B})$, are the
variable nodes; $\psi_\mathrm{o}$ and $\psi_\mathrm{e}$ are the factor nodes that indicates the compatibility associated with the variable nodes; The likelihood probabilities $\Pr(y_{2n-1}^\mathrm{R} \vert x^{n,n-1})$ and
$\Pr(y_{2n}^\mathrm{R} \vert x^{n,n})$ are the evidences from observation
$Y^\mathrm{R}$.}\label{fig6}
\end{figure}

We use a BP algorithm to compute soft information of $\Pr (x_n^\mathrm{A}
,x_n^\mathrm{B} \vert Y^\mathrm{R})$ from the $2N+1$ samples. The associated
Tanner graph is shown in Fig. \ref{fig6}. In the Tanner
graph $x^{i,j}\overset{\Delta}{=} (x_i^\mathrm{A} ,x_j^\mathrm{B} )$ are the
variable node; $\psi_\mathrm{o}$ and $\psi_\mathrm{e}$ are the compatibility
functions associated with the factor nodes. The compatibility functions
model the correlation between two adjacent symbols and are defined as
\begin{subequations}
\begin{align}
\label{eq191}
    \psi_\mathrm{o}\left(x^{n,n-1},x^{n,n}\right)&= \begin{cases}
        1 & \text{if the values of } x_n^\mathrm{A} \text{ in } x^{n,n-1} \text{ and
}x^{n,n}\text{ are equal}\\
        0 & \text{otherwise }
        \end{cases}\\
\label{eq192}
    \psi_\mathrm{e} \left(x^{n,n},x^{n+1,n}\right)&= \begin{cases}
        1 & \text{if the values of }x_n^\mathrm{B} \text{ in }x^{n,n}\text{ and }x^{n+1,n}\text{ are equal}\\
        0 & \text{otherwise }
        \end{cases}
\end{align}
\end{subequations}

The likelihood probabilities $\Pr(y_{2n-1}^\mathrm{R} \vert x^{n,n-1})$ and
$\Pr(y_{2n}^\mathrm{R} \vert x^{n,n})$ are the evidences from observation
$Y^\mathrm{R}$. The computation of these evidences is given by
\begin{subequations}
\begin{align}
\label{eq201}
    \Pr\left(\left.y_{2n-1}^\mathrm{R} \right|x^{n,n-1}\right)&=&\Pr
\left(\left.y_{2n-1}^\mathrm{R}\right| x_n^\mathrm{A} ,x_{n-1}^\mathrm{B}\right) &=&\frac{1}{\sqrt{2\pi\sigma^2/\tau}} \exp\left\{
{-\frac{\left| {y_{2n-1}^\mathrm{R} -x_n^\mathrm{A} -x_{n-1}^\mathrm{B} }
\right|^2}{{2\sigma ^2}/\tau }} \right\}\\
\intertext{and}
\label{eq202}
    \Pr \left( {\left. {y_{2n}^\mathrm{R} } \right|x^{n,n}} \right)&=&\Pr \left(
{\left. {y_{2n}^\mathrm{R} } \right|x_n^\mathrm{A} ,x_n^\mathrm{B} }
\right)&=&\frac{1}{\sqrt {{2\pi \sigma ^2}/(1-\tau)}}\exp \left\{ {-\frac{\left| {y_{2n}^\mathrm{R}
-x_n^\mathrm{A} -x_n^\mathrm{B} } \right|^2}{2\sigma ^2/(1-\tau)}} \right\}.
\end{align}
\end{subequations}

Given the evidences computed in (\ref{eq201}) and (\ref{eq202}), the message update equations can be
derived using the standard sum-product formula of BP
\cite{kschischang2001factor}. Note that the Tanner graph has a tree
structure. This means that the BP algorithm can compute the exact APP of
$x^{n,n}$ and $x^{n,n-1}$ for $n=1,\cdots,N$. Furthermore, the solution can
be found by passing the messages only once in each direction of the Tanner
graph. Although we can compute the APP of all variable nodes, we only use
the APP of $\Pr(x_n^\mathrm{A} ,x_n^\mathrm{B} \vert Y^\mathrm{R})$ for further
decoding.

\subsection{Codeword-Realignment Layer: Countering Integral Symbol Misalignment}
\label{subsec:codeword}
Since the fractional part of symbol misalignment has been removed in the
symbol-realignment layer, here we only consider the integral part of symbol
misalignment in this subsection. Recall that in Section
\ref{sec:synchronous} we used (\ref{eq14}) to compute the message $\gamma(\bar{c}_k)$. Equation (\ref{eq14}) requires that the modulated symbols of end
nodes A and B are symbol-by-symbol aligned (i.e., $x_n^\mathrm{A}$ must align
with $x_n^\mathrm{B}$). However, with integral symbol misalignment $\tau
_\mathrm{I} $, $x_n^\mathrm{A}$ will be aligned with $x_{n-\tau_\mathrm{I}
}^\mathrm{B}$ therefore the algorithm proposed in Section \ref{sec:synchronous} becomes invalid.

The codeword-realignment layer addresses this challenge using a specially
designed interleaver/deinterleaver at the end/relay nodes. At the end nodes,
we use the same block interleaver with $R$ rows and $M/R$ columns, where $r=1/R$ is
the code rate and $M$ is the number of bits in the codeword. To interleave,
the channel-coded bits are filled into the interleaver column-wise , and
read out row-wise. Then the interleaved packets are modulated and
transmitted simultaneously to the relay. Upon receiving the overlapped
signal (with symbol misalignment), the relay first deals with the fractional
symbol misalignment with the algorithm proposed in Section
\ref{subsec:symbol}. Then the relay uses the same block deinterleaver
to deinterleave the received signal.

Let us consider an example with code rate $1/3$ convolutional code, BPSK
modulation, and integral symbol misalignment $\tau=2$ (in this subsection,
we only consider the integral part of $\tau$). As specified in Section \ref{sec:system}, the channel-coded packets of node A and node B
are $C^\mathrm{A}$ and $C^\mathrm{B}$, respectively. Then the channel-coded
packet is bit-interleaved with a block interleaver with $M/R$ rows and $R$
columns. The interleaved packets $\tilde{C}^i,\,i\in {\mathrm{A,B}}$ are BPSK modulated to produce the transmitted signal
\begin{equation}
\label{eq21}
X^i=\left(x_{1,1}^i ,x_{2,1}^i ,\cdots ,x_{K,1}^i ,\;\;x_{1,2}^i ,\cdots
,x_{K,2}^i ,\;\;x_{1,3}^i ,\cdots ,x_{K,3}^i\right),\  i\in \left\{
\mathrm{A,B}\right\}
\end{equation}
where $x_{k,l}^i =1-2c_{k,l}^i$, and $c_{k,l}^i $is defined in (\ref{eq2}). The
received signal samples will be the superposition of the following two
sequences
\begin{equation}
\label{eq22}
\begin{matrix}
x_{1,1}^\mathrm{A}& x_{2,1}^\mathrm{A}& x_{3,1}^\mathrm{A}& x_{4,1}^\mathrm{A}& x_{5,1}^\mathrm{A}& \cdots& x_{K,3}^\mathrm{A}& &\\
&&&&+&&&&\\
& &x_{1,1}^\mathrm{B}&x_{2,1}^\mathrm{B}&x_{3,1}^\mathrm{B}&\cdots &x_{K-2,3}^\mathrm{B}&x_{K-1,3}^\mathrm{B}&x_{K,3}^\mathrm{B}.
 \end{matrix}
\end{equation}

The relay first aligns the unoverlapped (clear) part of signal:
$x_{1,1}^\mathrm{A} x_{2,1}^\mathrm{A}$ at the head and $x_{K-1,3}^\mathrm{B}
x_{K,3}^\mathrm{B}$ at the tail. Then the relay deinterleaves this packet to
restore node A's transmission order. After deinterleaving, the received
packet becomes the superposition of the following sequences
\begin{equation}
\label{eq23}
\begin{matrix}
x_{1,1}^\mathrm{A}& x_{1,2}^\mathrm{A}& x_{1,3}^\mathrm{A}& x_{2,1}^\mathrm{A}& x_{2,2}^\mathrm{A}& x_{2,3}^\mathrm{A}& \cdots &x_{K,1}^\mathrm{A}& x_{K,2}^\mathrm{A}& x_{K,3}^\mathrm{A}\\
&&&&&+&&&&\\
x_{K-1,3}^\mathrm{B}& x_{K-1,1}^\mathrm{B}& x_{K-1,2}^\mathrm{B}& x_{K,3}^\mathrm{B}&
x_{K,1}^\mathrm{B}& x_{K,2}^\mathrm{B}& \cdots  &x_{K-2,1}^\mathrm{B}& x_{K-2,2}^\mathrm{B}& x_{K-2,3}^\mathrm{B}.
\end{matrix}
\end{equation}

The signal in (\ref{eq23}) is equivalent to the superposition of modulated signals
of $C^\mathrm{A}$ and $C_{(6)}^\mathrm{B} $, where $C_{(6)}^\mathrm{B} $ denotes
the 6 bits right circular-shifted version of $C^\mathrm{B}$. Since tail biting
convolutional code with code rate $1/R$ is quasi-cyclic with period $R$ ,
$C_{(\tau R)}^\mathrm{B}$, the $\tau R$ bit circular-shifted version of
$C^\mathrm{B}$, is also a valid codeword \cite{ma1986tail,esmaeili1998link,solomon1979connection}. Hence we can apply the Jt-CNC
decoding algorithm proposed in Section \ref{sec:synchronous}.
Furthermore, as we prove in the Appendix, a very good property of
convolutional code is: the source packet corresponding to $C_{(\tau
R)}^\mathrm{B} $ is $U_{(\tau )}^\mathrm{B} $, the $\tau$-bit right
circular-shifted version of node B's source packet $U^\mathrm{B}$.

As a result, in the presence of integral symbol misalignment, our
bit-optimal decoding algorithm will output $U^\mathrm{R}=U^\mathrm{A}\oplus
U_{(\tau )}^\mathrm{B}$. However, node A (B) can still restore the
information of node B (A). Node A can first XOR $U^\mathrm{R}$ with its own
packet to obtain $U_{(\tau )}^\mathrm{B} $, then left shift it $\tau $ bits to
restore $U^\mathrm{B} $. Node B can first right shift its own packet to
obtain $U_{(\tau )}^\mathrm{B} $ and then XOR it with $U^\mathrm{R}$ to obtain
$U^\mathrm{A}$.

\subsection{ Asynchronous PNC with Linear Cyclic Codes}
\label{subsec:mylabel1}
In our discussion above, the quasi-cyclic property of convolutional code
plays a key role in countering asynchrony. Can other codes that have this
property be used to tackle asynchrony? To answer this question, we propose a
more general scheme that uses linear cyclic codes to cope with
larger-than-one symbol misalignment.

Let ${\mathcal C}(\cdot)$ and ${\mathcal C}^{-1}(\cdot)$ denote the encoding
function and decoding function of a particular linear cyclic code (e.g., BCH
code), respectively. Then the encoding process in the end nodes is
$C^i={\mathcal C}(U^i),\,i\in{\mathrm{A,B}}$. To
ease presentation, we assume BPSK modulation and a symbol misalignment of
$\tau$. The received signal at the relay is the overlap of the following
two signals:
\begin{equation}
\label{eq24}
\begin{matrix}
x_1^\mathrm{A}& \cdots& x_\tau^\mathrm{A}& x_{\tau +1}^\mathrm{A}& \cdots\cdots &x_N^\mathrm{A}&&& \\
&&&x_1^\mathrm{B}& x_2^\mathrm{B}\cdots & x_{N-\tau }^\mathrm{B}& x_{N-\tau+1}^\mathrm{B}& \cdots & x_N^\mathrm{B}.
\end{matrix}
\end{equation}

Upon receiving the overlapped signal, the relay first aligns the last $\tau
$ symbols with the first $\tau $ symbols to obtain a new overlapped signal
\begin{equation}
\label{eq25}
\begin{matrix}
x_1^\mathrm{A}& x_2^\mathrm{A}& \cdots & x_\tau^\mathrm{A}&
x_{\tau +1}^\mathrm{A}& \cdots & \cdots & x_N^\mathrm{A}\\
x_{N-\tau +1}^\mathrm{B}& \cdots & \cdots &x_N^\mathrm{B} &x_1^\mathrm{B} &x_2^\mathrm{B} &\cdots & x_{N-\tau }^\mathrm{B}.
\end{matrix}
\end{equation}

The result in (\ref{eq25}) is actually the signal $X^\mathrm{A}+X_{(\tau)}^\mathrm{B}
$, where $X_{(\tau )}^\mathrm{B}$ is the $\tau$-symbol right
circular-shifted version of node B's signal. Then the relay can map the
signal of (\ref{eq25}) to $C^\mathrm{A}\oplus C_{(\tau )}^\mathrm{B} $, where $C_{(\tau
)}^\mathrm{B}$ is the $\tau$-bit right circular-shifted version of
$C^\mathrm{B}$. Note that $C_{(\tau)}^\mathrm{B}$ is also a valid codeword
due to the property of cyclic code. We assume the source packet
corresponding to $C_{(\tau )}^\mathrm{B} $ is $\tilde{U}^\mathrm{B}$ such
that $\tilde{U}^\mathrm{B}={\mathcal C}^{\mathrm{-}1}(C_{(\tau )}^\mathrm{B})$. Because the XOR operator preserves the linearity of codes, the
relay first decode the XORed packet by
\begin{eqnarray}
\label{eq26}
U^\mathrm{R} &=& \mathcal{C}^{-1}\left(C^\mathrm{A}\oplus C_{(\tau )}^\mathrm{B}\right) \nonumber \\
 &=&\mathcal{C}^{-1}\left(C^\mathrm{A}\right)\oplus \mathcal{C}^{-1}\left(C_{(\tau)}^\mathrm{B}\right) \nonumber \\
 &=&U^\mathrm{A}\oplus \tilde{U}^\mathrm{B}
\end{eqnarray}
and then broadcasts this packet to both the end nodes. After decoding
$U^\mathrm{R}$ node A first XORs $U^\mathrm{R}$ with its own information
$U^\mathrm{A}$ to obtain $\tilde{U}^\mathrm{B} $; then node A re-encodes
$\tilde{U}^\mathrm{B}$ to obtain $C_{(\tau)}^\mathrm{B} ={\mathcal C}(
\tilde{U}^\mathrm{B})$, and left circular-shifts $C_{(\tau)}^\mathrm{B}$ to obtain $C^\mathrm{B}$; finally from $C^\mathrm{B}$ node A
can decode $U^\mathrm{B}$. For node B, it first right circular-shifts its
codeword $C^\mathrm{B}$ to produce $C_{(\tau)}^\mathrm{B}$ and decodes
$C_{(\tau )}^\mathrm{B} $ to obtain $\tilde{U}^\mathrm{B}$; then it XORs
$U^\mathrm{R}$ with $\tilde{U}^\mathrm{B}$ to obtain $U^\mathrm{A}$.

\section{Numerical Results}\label{sec:numerical}

We evaluate the performance of the proposed PNC decoding framework under AWGN channel by
extensive simulation. First, we compare the BER performances of Jt-CNC,
XOR-CD Viterbi, and full-state Viterbi algorithms in synchronous PNC.
Second, we demonstrate the effect of phase offset on our Jt-CNC decoder.
Third, we show the performance of Jt-CNC algorithm in the presence of symbol
and phase asynchrony. Furthermore, we implement the three algorithms in a
practical PNC system built on USRP software radio platform, and test them in
real indoor environment.

\subsection{BER Performance Comparison}
We compare the BER performances of Jt-CNC, XOR-CD Viterbi (XOR-CDV), and
full-state Viterbi (FSV) \lsc{in} synchronous PNC. The XOR-CD Viterbi algorithm
and full-state Viterbi algorithm were introduced in Section \ref{sec:related}. In the simulations, we adopt convolutional codes of two different code rates: code rate $1/2$ $(5, 7)$ code and code rate $1/3$ $(13, 15, 17)$ code. We first consider BPSK modulation, assuming AWGN channel.
\begin{figure}[htb]
\centering
  \includegraphics[width=12cm]{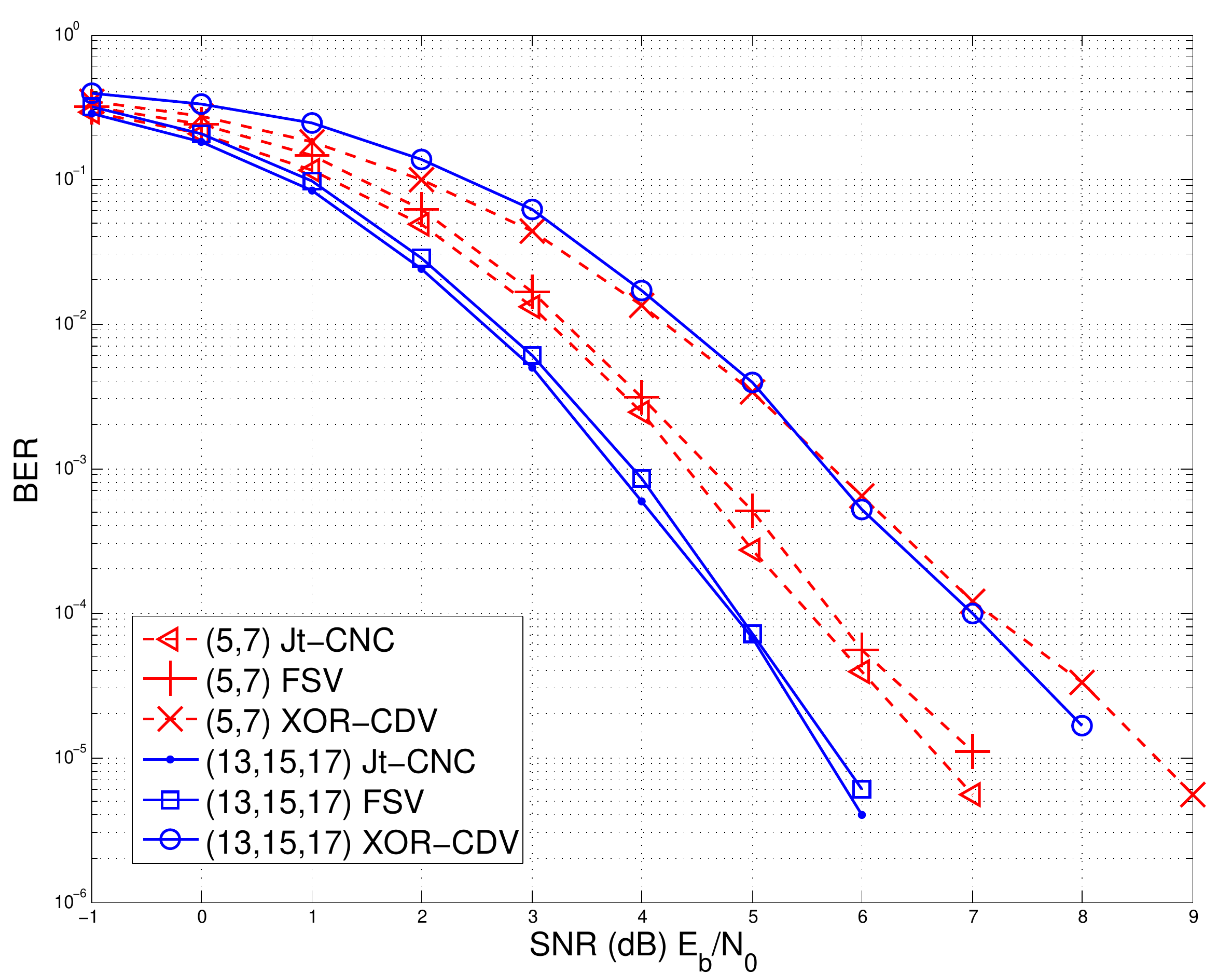}
  \caption{BER performance of Jt-CNC, XOR-CD Viterbi (XOR-CDV), and full-state Viterbi (FSV) algorithms with PNC. The channel codes are $(5, 7)$ and $(13, 15, 17)$ convolutional code. We use BPSK modulation and assume AWGN channel.}\label{fig7}
\end{figure}

We plot the BER curve of the full-state Viterbi algorithm as a benchmark for
the reduced-state Viterbi algorithm in \cite{to2010convolutional}. In our
attempt to replicate the reduced-state Viterbi algorithm, we cannot get the same simulation results in \cite{to2010convolutional} even though we follow the exact specification as described in the paper\footnote{We believe that there are errors in equation (12) and Fig. 3 in \cite{to2010convolutional}. We suspect that in \cite{to2010convolutional}, the SNR was not normalized correctly. Our attempt to contact the authors of \cite{to2010convolutional} by email received no reply.}. Our simulation results are somewhat
better than those presented in \cite{to2010convolutional}. To avoid misrepresenting their results, here we just compare the results of full-state \lsc{Viterbi with Jt-CNC. In \cite{to2010convolutional}, a performance gap of \unit[2]{dB} was observed between the reduced-state Viterbi and full-state Viterbi.} As shown in Fig. \ref{fig7}, Jt-CNC has better BER performance than full-state Viterbi.
If the gap between full-state Viterbi and reduced-state Viterbi is \unit[2]{dB}, then
the gap between Jt-CNC and reduced-state Viterbi is at least \unit[2]{dB}.

Fig. \ref{fig7} also shows
that Jt-CNC outperforms XOR-CD Viterbi by \unit[2]{dB} for both rate $1/2$
and $1/3$ convolutional codes. As described previously, XOR-CD loses
information in the XOR-mapping, hence this \unit[2]{dB} gap is as expected.

\begin{figure*}[!t]
\centerline{\subfloat[without random-phase precoding]{\includegraphics[height=3.0in]{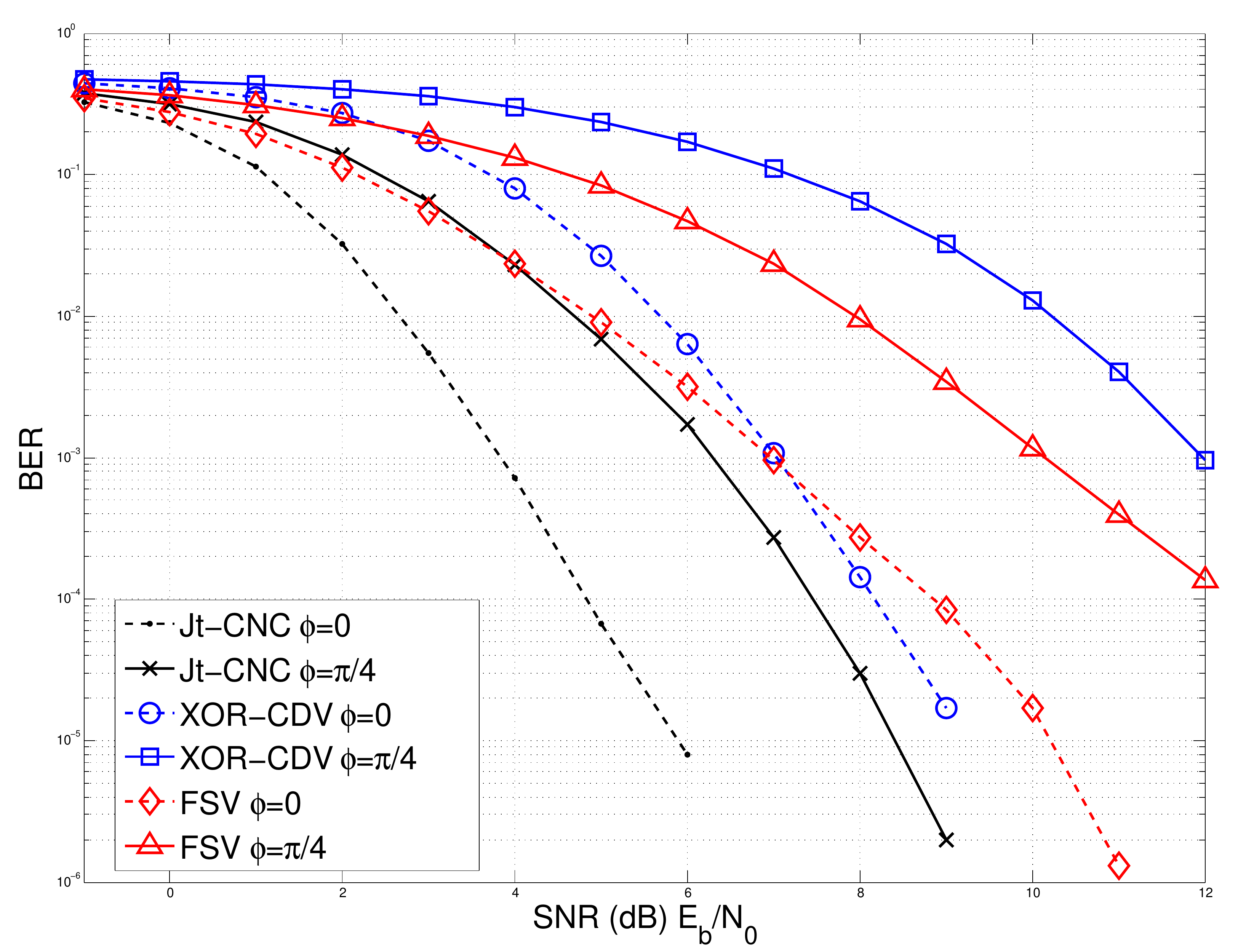}%
\label{fig:fixed_phase}}
\hfil
\subfloat[with random-phase precoding]{\includegraphics[height=3.0in]{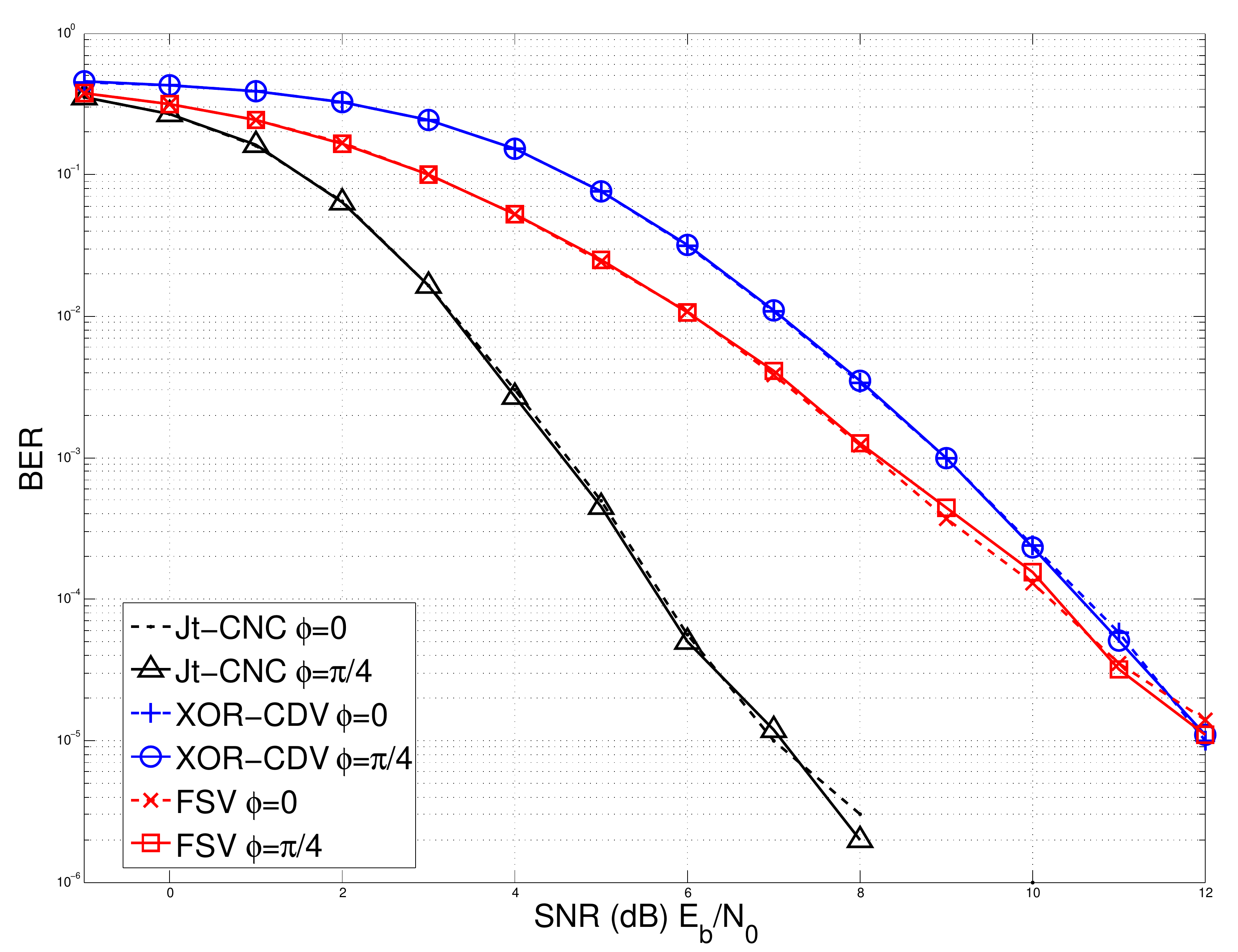}%
\label{fig:random_phase}}}
\caption{Effects of phase offset on Jt-CNC, XOR-CD Viterbi (XOR-CDV), and full-state Viterbi (FSV). QPSK modulation and $(13, 15, 17)$ convolutional code are used in the simulation. We assume the symbols are aligned and the relative phase offset is $\pi/4$. In (a), both nodes transmit their signals directly; in (b), node B precodes its transmit signal with a pseudo-random phase sequence.}
\label{fig:phase_offset}
\end{figure*}

\subsection{Effects of Phase Offset}
\label{subsec:effects}
We next evaluate the effect of phase offset on Jt-CNC decoder assuming QPSK
modulation (higher order QAM can also be used)---phase offset does not
present a challenge to BPSK systems \cite{lu2012asynchronous,zhang2009channel}.
First, we compare the BER performances of the aforementioned three decoding algorithms with phase offset $\phi{=}0$
(phase synchronous) and $\phi{=}\pi/4$ (worst case for QPSK) \cite{lu2012asynchronous}.

As shown in Fig. \ref{fig:fixed_phase}, when the phase offset is $\pi/4$, the BER performances of Jt-CNC,
FSV, and XOR-CDV are degraded by \unit[2]{dB}, \unit[3]{dB}, and \unit[5]{dB}, respectively. The severe phase penalty is due to the poor confidence of the messages as calculated in (\ref{eq14}) when the phase offset is $\pi/4$.
One method to improve the confidence is to make the phase offset random so that the symbols with small phase offset can help the symbols with large phase
offset during the BP process.

As shown in Fig. \ref{fig:phase_offset}, phase offset degrades the performances of all the three algorithms.  Jt-CNC is more resistant to phase penalty because the joint
processing makes better use of the soft information. We note the performance
of Jt-CNC under the worst phase offset of $\pi/4$ is still better than the
performance of XOR-CDV with no phase offset.

To improve our system's resilience against phase
offset, we adopt the \textit{random-phase precoding} at the transmitter of one end node. Specifically, node B rotates the phase of its
 transmitted signal with a pseudo-random phase sequence $\Phi^{\mathrm{B}}{=}(\phi^{\mathrm{B}}_1, \cdots, \phi^{\mathrm{B}}_N)$
 where $\phi^{\mathrm{B}}_n$ is randomly chosen from zero to $\pi/4$. We assume that this pseudo-random phase sequence is known at the relay so that it can incorporate this knowledge into the decoding process. As shown in Fig. \ref{fig:random_phase} with the random-phase precoding algorithm, the phase penalty is reduced
 to \unit[1]{dB}, \unit[1]{dB}, and \unit[3]{dB} for Jt-CNC, FSV, and XOR-CDV, respectively.

\subsection{Effects of Symbol Misalignment}
\label{subsec:mylabel2}
A major advantage of the proposed decoding framework is that it can deal
with general symbol misalignment. We evaluate the performance of Jt-CNC
under varying degrees of symbol misalignment and phase offset. In the
simulation, both end nodes transmit 1000-bit source packets (corresponding
to 1500 QPSK symbols for channel code rate of $1/3$).

\begin{figure}[htb]
\centering
  \includegraphics[width=12cm]{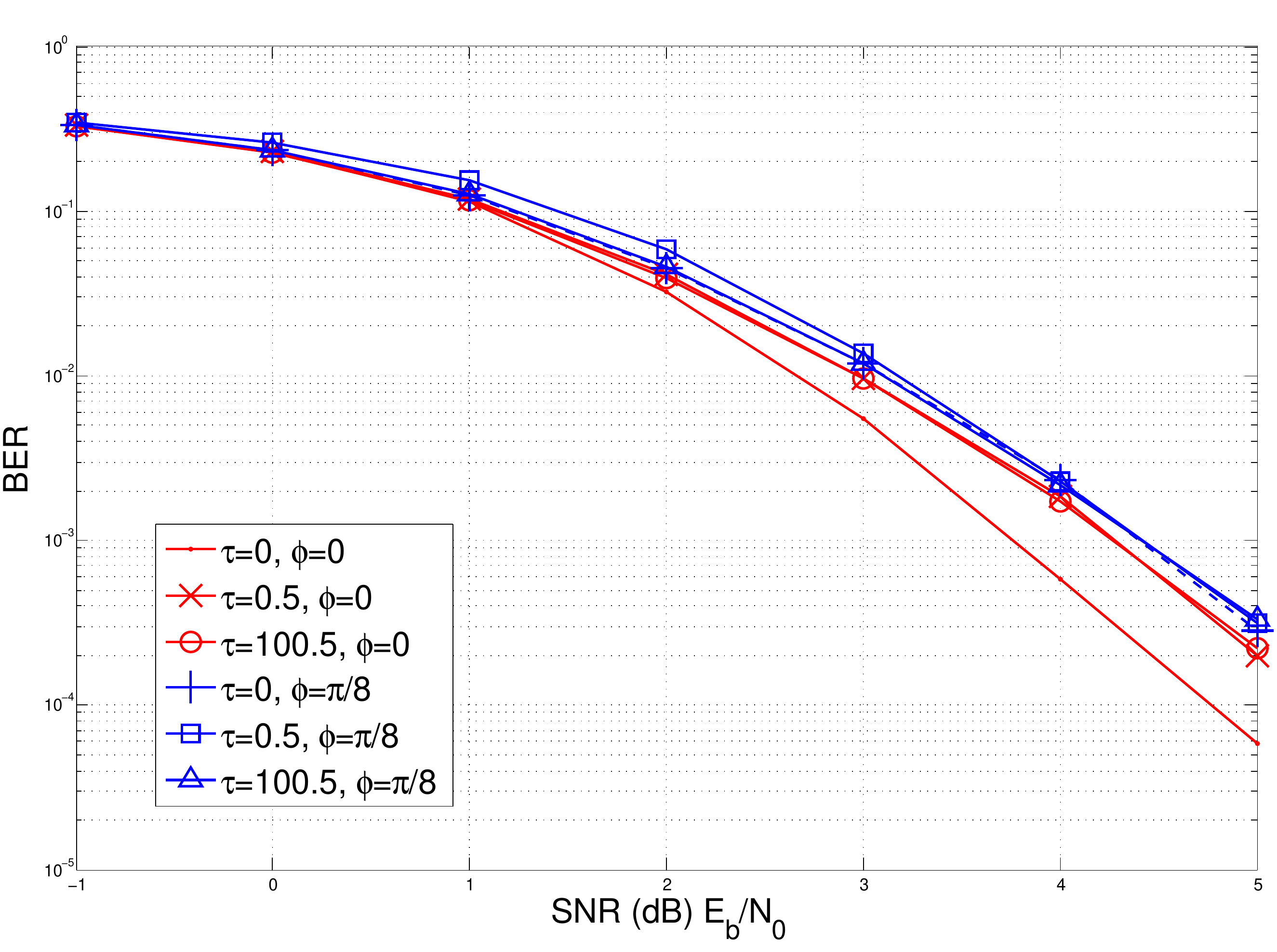}
  \caption{BER performance of Jt-CNC decoder under general symbol misalignment, with $(13, 15, 17)$ convolutional code and QPSK modulation.}\label{fig:symbol_asyn}
\end{figure}

From Fig. \ref{fig:symbol_asyn} we see that although the fractional
symbol misalignment (the curve with $\tau =0.5,\phi =0)$ degrades the
BER performance by 0.5dB, the integral symbol misalignment (the
curve with $\tau {=}100.5,\phi {=}0)$ improves the BER performance
slightly. That is because when there are integral symbol misalignments, the
head and tail of the signals are non-overlapping and thus yield cleaner
information without the mutual interference.

\subsection{Software Radio Experiment}
\label{subsec:software}
To evaluate the proposed algorithm in a real communication system, we
implemented a PNC system using USRP N210,
embedded with the three decoding algorithms. The PNC system adopts OFDM
modulation with 1MHz bandwidth and \unit[2.48]{GHz} carrier frequency. We use the $(5, 7)$ convolutional code and follow the frame format design in \cite{lu2013real}. We conduct our experiments in the indoor
office environment and evaluated the
BER performance of Jt-CNC, XOR-CD, and full-state Viterbi algorithms under different
SNRs. In the experiment, we balanced the powers of the end nodes and let
both nodes transmit 500 frames to the relay. Each frame consisted of 204
OFDM symbols (4 symbols of preambles and 200 symbols of data).

\begin{figure}[htb]
\centering
  \includegraphics[width=12cm]{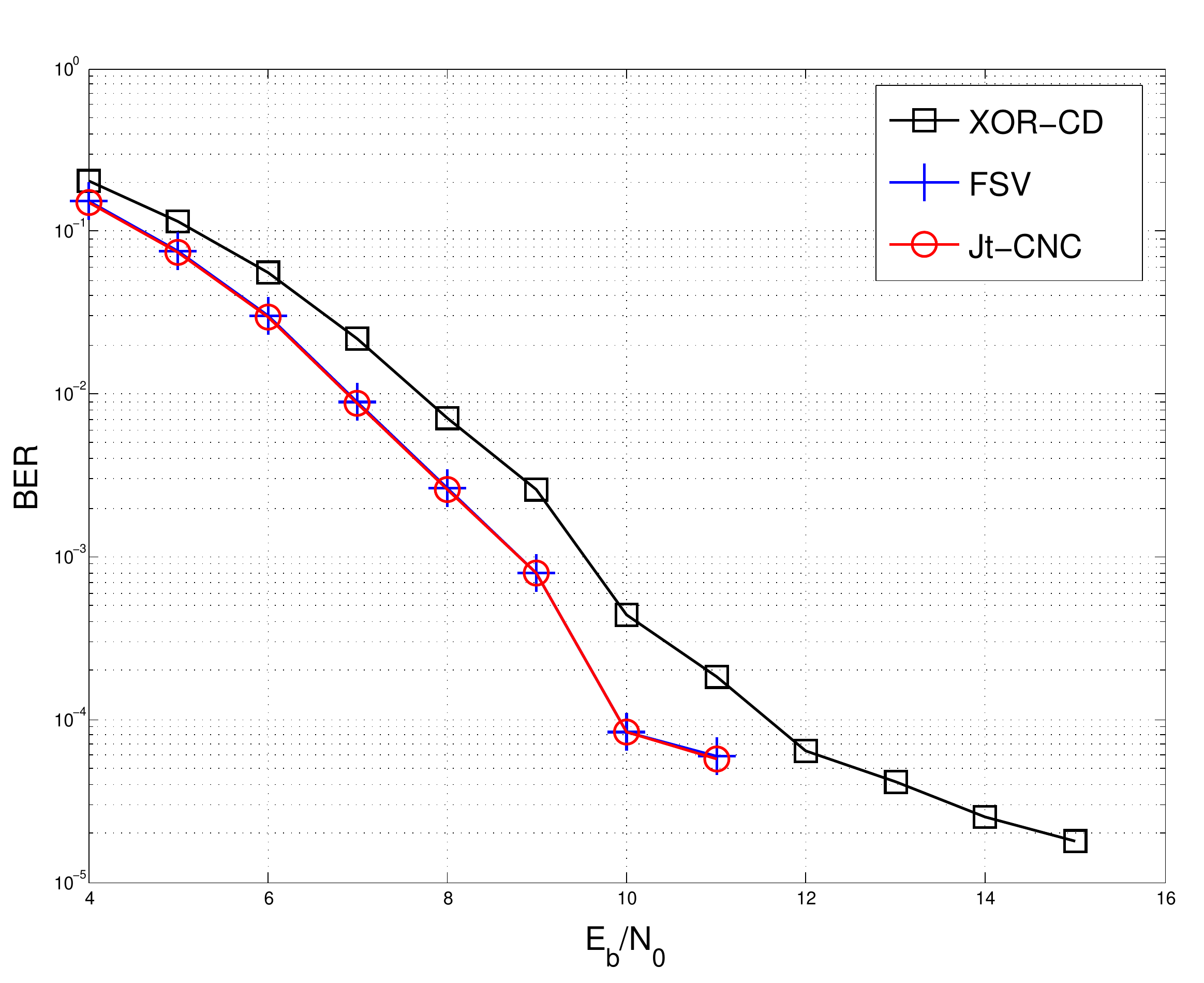}
  \caption{BER performance of Jt-CNC, FSV, and XOR-CD in an indoor environment. We tested the three algorithms on a practical PNC system implemented on USRP N210. The PNC system adopts OFDM modulation and $(5, 7)$ convolutional code.}\label{fig:exp}
\end{figure}

As shown in Fig. \ref{fig:exp} the BER performances of
Jt-CNC and full-state Viterbi are nearly the same in real indoor environment. XOR-CD, however, is worse
by about \unit[2]{dB} at $10^{-4}$ BER. Compared with the simulation results in
Fig. \ref{fig7}, the BER performance of all the three
algorithms in the real system are degraded by \unit[4]{dB} due to imperfections in
the real systems, such as imperfect channel estimation, carrier-frequency
offsets, and frequency-selective channels. We also note that XOR-CD has an
error floor even in the high SNR regime while the other two algorithms do
not.

\section{Conclusion}
\label{sec:conclusion}
We have proposed a three-layer decoding framework for asynchronous
convolutional-coded PNC systems. This framework can deal with general
(integral plus fractional) symbol misalignment in convolutional-coded PNC
systems. Furthermore, we design a Jt-CNC algorithm to achieve the
BER-optimal decoding of convolutional code in synchronous PNC. For
asynchronous PNC, the performance degradation is within \unit[1]{dB}. Building on the study of
convolutional codes, we further generalize the Jt-CNC decoding algorithm to
all cyclic codes, providing a new angle to counter symbol asynchrony. Simulation shows that our Jt-CNC algorithm outperforms the previous decoding
algorithm (XOR-CD, reduced-state Viterbi) by \unit[2]{dB}. \yq{With random-phase precoding, the proposed Jt-CNC algorithm is more resilient to phase offset than XOR-CD and full-state Viterbi.} Importantly, we have implemented  the proposed Jt-CNC decoder in a real PNC system built on software radio platform. Our experiment shows that the Jt-CNC decoder works well in practice.

\appendix
\begin{theorem}\label{theorem1}
For a tail biting convolutional code with code
rate $1/R,\;R \in \mathbf{N}^+$, let $U$ denote the source packet of the channel-coded packet $C$, and $C_{(kR)}$ denote the $kR$-bit right
circular-shifted version of $C$. The source packet corresponding to $C_{(kR)}$ is $U_{(k)}$, the $k$-bit right circular-shifted version of $U$.
\end{theorem}
\begin{figure}[h!]
\centering
  \includegraphics[width=6cm]{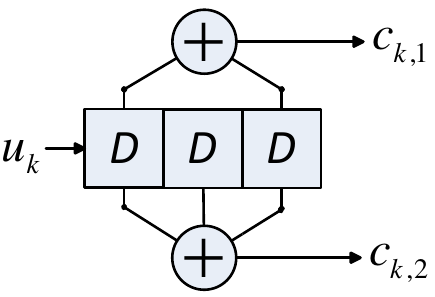}
  \caption{Convolutional encoder of $(5, 7)$ convolutional code. $u_k$ is the input source bit at time $k$; $c_{k,1}$ and $c_{k,2}$ are the first and second output bit of the encoder at time $k$, respectively.}\label{fig:57_enc}
\end{figure}

\begin{IEEEproof}
Let $m$ denote the memory length of this
convolutional encoder. The generator matrix of the convolutional code is
\begin{equation}
\rm{\bf G}=\begin{bmatrix}
 {{\rm{\bf g}}_0}& {{\rm {\bf g}}_1}& {{\rm {\bf g}}_2 }
& \cdots & {{\rm {\bf g}}_m }  &  &  &\\
& {{\rm {\bf g}}_0 }& {{\rm {\bf g}}_1 }& \cdots
 & {{\rm {\bf g}}_{m-1} }& {{\rm {\bf g}}_m } & & \\
&  & \ddots& & & & \ddots & \\
& & & {{\rm {\bf g}}_0 } & {{\rm {\bf g}}_1 }& {{\rm {\bf g}}_2 } & \cdots & {{\rm {\bf g}}_m }\\
 {{\rm {\bf g}}_m } & & & & {{\rm {\bf g}}_0 }& {{\rm {\bf g}}_1 } & \cdots & {{\rm {\bf g}}_{m-1} }\\
 {{\rm {\bf g}}_{m-1} } \hfill & {{\rm {\bf g}}_m } \hfill & \hfill & \hfill
& & {{\rm {\bf g}}_0 } & \cdots & {\rm {\bf g}}_{m-2} \\
 \vdots & & \ddots & & & & \ddots& \\
 {{\rm {\bf g}}_1 } & {{\rm {\bf g}}_2 } & \cdots &{{\rm {\bf g}}_m } & & & & {{\rm {\bf g}}_0 } \\
\end{bmatrix}
\end{equation}
where ${\rm{\bf g}}_\mathrm{b} =[
 {{\rm {\bf g}}_0 }\ {{\rm {\bf g}}_1 } \ \cdots \
{{\rm {\bf g}}_m }]$ is the basis generator matrix of the convolutional
code; each entry ${\rm {\bf g}}_i$ is an $R$-bit vector
\begin{equation}
\mathbf{g}_i =\begin{bmatrix}
 g_i^{(1)} & g_i^{(2)} & \cdots & g_i^{(R)}
\end{bmatrix}
\end{equation}
where $g_i^{(r)}$ is equal to 1 or 0, corresponding to whether
the $i$th stage of the shift register contributes (connects) to the $r$th output.
Therefore, the basis generator matrix ${\rm{\bf g}}_{\rm b} $ can be
regarded as the ``impulse response'' of the convolutional encoder. For
example, the basis generator matrix of $(5,\,7)$ convolutional code shown in
Fig. \ref{fig:57_enc}  is $[11\ 01\ 11]$. The encoding process is simply
\begin{equation}
C=U{\rm {\bf G}}.
\end{equation}

The right circular-shifted codeword can be represented by
\begin{equation}
C_{(kR)} =U{\rm {\bf G}}_{(k)}
\end{equation}
where ${\rm {\bf G}}_{(k)} $ is obtained by right circular-shift matrix
${\rm {\bf G}}$ by $k \times R$ columns. Since ${\rm {\bf G}}$ is a circulant matrix,
we have
\begin{equation}
\label{}
C_{(kR)} =U{\rm {\bf G}}_{(k)} =U_{(k)} {\rm {\bf G}}.
\end{equation}

Therefore the source packet of $C_{(kR)} $ is $U_{(k)} $, the $k$-bit
circular-shifted version of $U$.
\end{IEEEproof}

\begin{remark}
Theorem \ref{theorem1} is also valid for the tail biting
convolutional code with a general code rate $L/R\;L,R \in \mathbf{N}^+$, but the resulting source
packet will be $U_{(kL)}$, the $kL$-bit right circular-shifted version of $U$.
The proof is the same except that the entry of the basis generator matrix
${\rm {\bf g}}_{\rm {\bf b}} $ is an $L \times R$ matrix:
\begin{equation}
\mathbf{g}_i = \begin{bmatrix}
g_{1,i}^{(1)} & g_{1,i}^{(2)} & \cdots & g_{1,i}^{(R)}\\
g_{2,i}^{(1)} & g_{2,i}^{(2)} & \cdots & g_{2,i}^{(R)}\\
\vdots & \vdots &  & \vdots \\
g_{L,i}^{(1)} & g_{L,i}^{(2)} & \cdots & g_{L,i}^{(R)} \\
\end{bmatrix}
\end{equation}
where $g_{l,i}^{(r)}$ is equal to 1 or 0, depending on whether
the $i$th stage of the shift register for the $l$th input contributes (connects)
to the $r$th output.
\end{remark}

\begin{theorem}\label{theorem2}
If the initial state and terminal state of a
convolutional code are the same, then the decoding output of $C_{(kR)}$,
the $kR$-bit right circular-shifted version of $C$, is the $k$-bit right
circular-shifted version of $U$.
\end{theorem}
\begin{IEEEproof}
The encoding and decoding process of a
convolutional code can be represented by the Tanner graph in
Fig. \ref{fig2}. Since the code has the same initial and
terminal state, we can merge the Tanner graph as shown in
Fig. \ref{fig:circle_tanner}. For a general convolutional code with
code rate $L/R$, the source message $\bar{u}_k$ is an $L$-bit tuple and the
coded message $\bar{c}_k $ is an $R$-bit tuple.

\begin{figure}[htb!]
\centering
  \includegraphics[width=10cm]{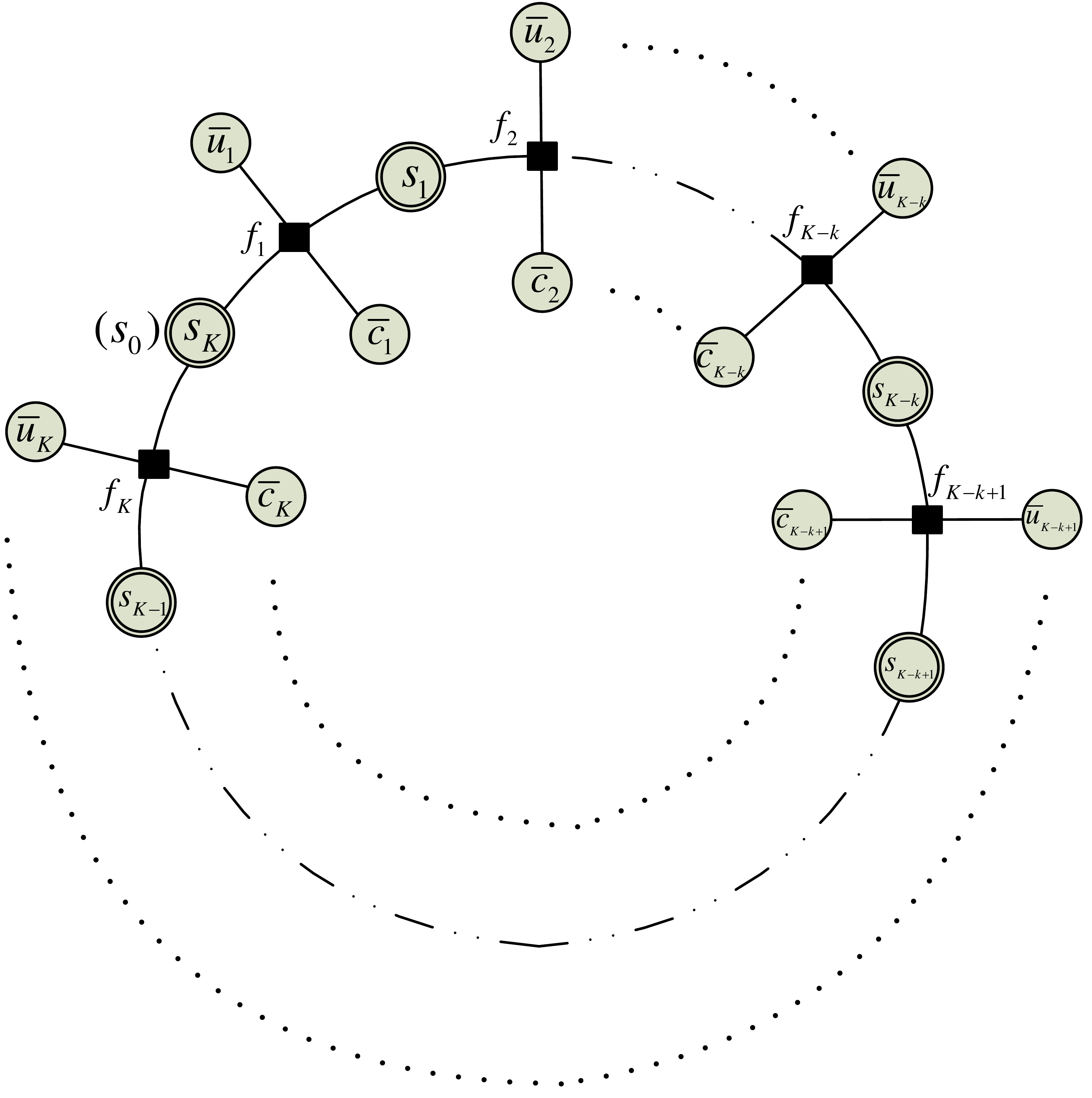}
  \caption{Tanner graph of a convolutional code that has the same initial state $S_0 $ and terminal state $S_K $. We merge the initial state and terminal state, hence the Tanner graph in Fig. \ref{fig2} becomes a ring.}\label{fig:circle_tanner}
\end{figure}

Let $C_{(kR)} =(\bar{c}_{K-k+1} ,\bar{c}_{K-k+2} ,\cdots ,\bar{c}_K
,\bar {c}_1 ,\cdots ,\bar{c}_{K-k})$ be the $kR$-bit right circular-shifted
version of codeword $C$. To decode $C_{(kR)}$, the decoding algorithm
starts with the first tuple $\bar{c}_{K-k+1}$ and ends with the last tuple
$\bar{c}_{K-k}$. Because the Tanner graph has a ring structure, the decode
output is $U_{(k)} =(\bar{u}_{K-k+1} ,\bar{u}_{K-k+2} ,\cdots ,\bar{u}_K ,\bar{u}_1 ,\cdots ,\bar{u}_{K-k})$, which is the $k$-bit right circular-shifted version of $U$.
\end{IEEEproof}

\begin{remark}
Zero tailing convolutional codes also have the
property in Theorem \ref{theorem2}, because zero tailing convolutional code has zero
initial state and zero terminal state.
\end{remark}

\begin{remark}
For a recursive convolutional code, we can
append zero-tailing bits to the input packet to make the terminal state zero
state. Then recursive convolutional codes can also be used with the proposed
Jt-CNC decoder.
\end{remark}

\section*{Acknowledgment}
The authors would like to thank...

\ifCLASSOPTIONcaptionsoff
  \newpage
\fi

%
\bibliographystyle{IEEEtran}
\bibliography{IEEEabrv,yq}

\begin{IEEEbiography}{xxxx}
Biography text here.
\end{IEEEbiography}

\begin{IEEEbiographynophoto}{xxxx}
Biography text here.
\end{IEEEbiographynophoto}


\begin{IEEEbiographynophoto}{xxxx}
Biography text here.
\end{IEEEbiographynophoto}




\end{document}